\documentclass[a4paper,fleqn,usenatbib,english]{mnras}
\usepackage[T1]{fontenc}
\usepackage{ae,aecompl}
\usepackage{amssymb}
\usepackage{amsmath}
\usepackage{babel}
\usepackage{graphicx}
\usepackage{amsmath}
\usepackage{amssymb}
\usepackage{subfig}
\usepackage{rotating}
\usepackage{longtable}
\usepackage{bigints}

\makeatletter

\newcommand{\rom}[2]{\ensuremath{#1_{\textrm{#2}}}}

\newcommand{\ebv}{\ensuremath{E(B\,\textrm{--}\,V)}}
\newcommand{\um}{$\mu$m}
\newcommand{\ium}{$\mu$m$^{-1}$}

\DeclareMathAlphabet\mathbfcal{OMS}{cmsy}{b}{n}

\title[Photo-z's, Fuzzy Templates, and SOMs. I.]{Deriving Photometric Redshifts using Fuzzy Archetypes and Self-Organizing Maps. I. Methodology}

\author[Speagle and Eisenstein]{
	Joshua S. Speagle$^{1,2}$\thanks{E-mail: jspeagle@cfa.harvard.edu}
	and Daniel J. Eisenstein$^{1}$
	\\
	$^{1}$Harvard University Department of Astronomy, 60 Garden St., MS 46, Cambridge, MA 02138, USA\\
	$^{2}$Kavli IPMU (WPI), UTIAS, The University of Tokyo, Kashiwanoha 5-1-5, Kashiwa, Chiba, Japan\\
}

\date{Accepted XXX. Received YYY; in original form ZZZ}

\pubyear{2015}

\begin{document}
\label{firstpage}
\pagerange{\pageref{firstpage}--\pageref{lastpage}}
\maketitle
	
\begin{abstract}
We propose a method to substantially increase the flexibility and power of template fitting-based photometric redshifts by transforming a large numbers of galaxy spectral templates into a corrresponding collection of ``fuzzy archetypes'' using a suitable set of perturbative priors designed to account for empirical variation in dust attenuation and emission line strengths. To bypass widely seperated degeneracies in parameter space (e.g., the redshift-reddening degeneracy), we train Self-Organizing Maps (SOMs) on a large ``model catalogs'' generated from appropriate Monte Carlo sampling of our fuzzy archetypes to cluster the predicted \textit{observables} in a topologically smooth fashion. Subsequent sampling over the SOM then allows full reconstruction of the relevant probability distribution functions (PDFs) using the associated set of inverse mappings from the SOM to the underlying model parameters. This combined approach enables the \textit{multi-modal} exploration of known variation among galaxy spectral energy distributions (SEDs) using large numbers of archetypes with minimal modeling assumptions. We demonstrate the power of this approach to recover full redshift PDFs using discrete Markov Chain Monte Carlo (MCMC) sampling methods combined with SOMs constructed from model catalogs based on LSST $ugrizY$ and \textit{Euclid} $YJH$ mock photometry.
\end{abstract}




\section{Introduction}
\label{sec:intro}

Cosmological studies must utilize not only the precise astrometric locations of large numbers of galaxies on the sky but also their associated redshifts ($z$) in order to study the expansion history of the universe, the growth of cosmic structure, and the nature of dark energy. Although spectroscopic redshifts (spec-z's) are often extremely precise, their time-intensive requirements guarantee that large-scale imaging surveys conducted by, e.g., Hyper-Suprime-Cam (HSC), \textit{Euclid}, Wide-Field Infrared Survey Telescope (\textit{WFIRST}), and the Large Synoptic Survey Telescope (LSST) will be unable to rely on them for large fractions of the faint galaxies that will populate their enormous samples. As a result, these and other existing surveys instead must rely on ``photometric redshifts'' (photo-z's) -- estimates of the associated redshift probability distribution function (PDF) $P(z)$ to a given object derived from imaging data -- to estimate the redshifts to the majority of these objects in a feasible amount of time.

A large number of techniques have been developed to derive photo-z's with varying levels of effectiveness \citep{hildebrandt+10,dahlen+13,sanchez+14}. These can broadly be sub-divided into two main categories: 
\begin{itemize}
	\item \textit{Template fitting} approaches, which rely on deriving a representative set of \textit{forward} mappings and their associated likelihoods through the exploration of a set of model parameters (including redshift) to color space based on a collection of observed and/or synthetic templates.
	\item \textit{Machine learning} approaches, which attempt to determine a representative set of \textit{inverse} mappings from color space (and possibly other associated observables) to redshift using an empirical ``training set'' of data.
\end{itemize}

In the last few years, the majority of development in the photo-z community has been centered on the increasingly sophisticated use of machine-learning techniques \citep[see, e.g.,][just to name a few]{carrascokindbrunner13,carrascokindbrunner14,bonnett15,hoyle15,almosallam+15,elliott+15}. This focus is in part due to two major outstanding issues with current template-fitting codes: 
\begin{enumerate}
	\item \textit{Computational expedience}: Many machine-learning architectures often can generate photo-z estimates significantly faster than their current template-fitting counterparts, a necessary feature for an algorithm that will eventually be applied to large numbers of objects.
	\item \textit{Modeling uncertainties}: Although the effectiveness of machine-learning approaches is limited by the quality of their training set, they are by and large free from the large systematic model uncertainties associated with dust and emission line contributions that tend to dominate template fitting-based photo-z codes today.
\end{enumerate}

However, while machine-learning methods possess many attractive features that make them well-posed to predicting photo-z's, they are unable to incorporate information about the physical processes involved. In particular, they are unable to make use of the fact that redshift evolution is a smooth and continuous process that affects the underlying galaxy SEDs in specific ways which are well understood, or that galaxies over the course of their lifetime evolve in a very particular manner through color space. While template-fitting approaches are fundamentally limited by the accuracy of our templates and associated modeling assumptions, they \textit{do} allow us to leverage our preexisting knowledge and physical intuition to better explore the mapping between color space and redshift.

We attempt to address some of the modeling deficiencies in current template-fitting methods in order to take advantage of these qualities and make them more competitive with modern machine-learning techniques. Our fundamental approach relies on promoting an underlying set of \textit{observed} galaxy spectra (i.e. ``archetypes'') to a corresponding set of ``fuzzy clouds'' in color space using small perturbations in dust extinction and emission line strengths in order to better approximate the complicated color-space manifold where galaxies live and evolve. These ``fuzzy templates'' enable us to take advantage of well-motivated modeling assumptions related to the photo-z problem (moving templates to different redshifts) without introducing the unnecessary nuisance parameters mentioned above that tend to plague traditional template-fitting approaches. This empirical incorporation of existing galaxy spectra is key, especially since the use of linear/non-negative combinations stellar population synthesis models \citep[from, e.g.,][]{bruzualcharlot03} or other atomic SEDs \citep[as used in, e.g., EAZY;][]{brammer+08} still have not been able to fully replicate observed galaxy spectra and/or colors \citep{cool+13,momcheva+15}.

To enable efficient exploration of the ensuing high-dimensional parameter space spanned by this large collection of archetypes, we use an unsupervised machine-learning algorithm known as a Self-Organizing Map \citep[SOM;][]{kohonen01} to construct a non-linear projection of a ``model catalog'' (a large number of Monte Carlo (MC) realizations of our fuzzy templates and a corresponding set of priors) onto a reduced-dimensional manifold. Once the model catalog has subsequently been ``mapped'' onto the reduced dimensional manifold spanned by the SOM, we can utilize approaches that sample ``on top of'' the SOM to estimate $P(z)$. This framework represents a computationally efficient approach to use observed spectra \textit{en masse} to derive photometric redshifts.

The paper is organized as follows. In \S\ref{sec:fuzzy}, we outline the procedure we use to create our fuzzy archetypes and the ``linearized'' scheme for utilizing them in a computationally efficient way. In \S\ref{sec:som}, we showcase how SOMs allow for smooth non-linear projections of both rest- and observed-frame model catalogs to well-localized clusters in lower dimensions. In \S\ref{sec:mcmc}, we demonstrate how a simplistic implementation of discrete Markov Chain Monte Carlo (MCMC) sampling over SOMs can efficiently explore large swaths of parameter space and recover $P(z)$. In \S\ref{sec:disc}, we discuss a variety of different ways to expand upon the simplistic framework outlined here both in concept and in practice. We summarize our results in \S\ref{sec:conc}.

Throughout this work, boldface font ($\mathbf{x}$, $\mathbf{\Theta}$) is used to represent vectors and matrices while normal font ($x$, $\Theta$) is used for singular variables.

\section{Exploring Parameter Space with Fuzzy Templates}
\label{sec:fuzzy}

\subsection{Generating Photometry}
\label{subsec:gen_phot}

To generate model photometry, most template-based codes begin with a set of ``basis'' galaxy templates $S_{\nu,\textrm{gal}}(\nu)$ \citep[see, e.g.,][]{coleman+80,kinney+96,polletta+07} modified with a set of emission line templates $S_{\nu,\textrm{lines}}(\nu)$ co-added according to a set of scaling relations taken from the literature \citep[e.g.,][]{kennicutt98,kennicuttevans12}. Rest-frame galaxy templates are created by superimposing a \textit{uniform screen} of galactic dust taken from a basis set of (normalized) dust templates $k_{\textrm{dust}}(\nu)$ parameterized by $\ebv$. The templates are then redshifted by (1+$z$), after which extinction from the intergalactic medium (IGM) $A_{\textrm{IGM}}(\nu|z)$ is superimposed to form the final observed-frame galaxy template. This final template is then convolved with the transmission of a particular set of filters (including atmospheric effects) $T(\nu)$ normalized to a source at constant flux density \citep[AB magnitude standard;][]{okegunn83} to generate the final corresponding set of model fluxes $\rom{\mathbf{F}}{model}$.

More formally,
\begin{equation}\label{eq:phot}
\rom{\mathbf{F}}{model}=\frac{\int_{\nu_z}  S_{\nu,\textrm{model}}(\nu) R_{\textrm{model}}(\nu) \mathbf{T}(\nu)\nu^{-1}d\nu}{\int_{\nu_z} \mathbf{T}(\nu) \nu^{-1} d\nu},
\end{equation}
where $\nu_z$ is the redshifted set of frequencies probed by the filters, the galaxy model
\begin{equation}
S_{\nu,\textrm{model}}(\nu)=\sum_{\textrm{gal}} c_{\textrm{gal}} \times  S_{\nu,\textrm{gal}}(\nu) + \sum_{\textrm{lines}} c_{\textrm{lines}} \times  S_{\nu,\textrm{lines}}(\nu)
\end{equation}
is the sum of a linear combination of galaxy and emission line templates, and the dust (i.e. reddening) model
\begin{equation}
-2.5\log \left(R_{\textrm{model}}(\nu)\right) = \sum_{\textrm{dust}}\rom{\ebv}{dust}\rom{k}{dust}(\nu) + \rom{A}{IGM}(\nu,z)
\end{equation}
is the sum of a linear combination of dust templates at a given $\ebv$ and reddening from the IGM whose form is given in \citet{madau95}.\footnote{While a number of codes allow for the exploration of limited linear combinations of galaxy and/or emission line templates during the fitting process \citep[e.g.,][]{blantonroweis07,brammer+08}, due to the non-linear effects of dust screens often only a single dust template is used to modify the underlying rest-frame galaxy model.}

\subsection{The ``Traditional'' Approach}
\label{subsec:phot_trad}

Due to the amount of time it takes to generate a single set of model fluxes using the process outlined above, most codes opt to use large \textit{pre-generated grids} of photometry generated from combinations of parameters sampled at a given granularity in each dimension rather than generate photometry ``on the fly''. The $P(z)$ for each object is then derived using the combined set of likelihoods computed at each grid point. As discussed in \citet{speagle+15} and references therein, this approach is reasonable given the grid can be constructed with sufficient resolution and explored relatively quickly using either efficiently pipelined ``brute force'' approaches or adaptive sampling techniques.

However, most of these grids are constructed using only a small collection of galaxy ($\lesssim$\,$30$) and dust ($\lesssim$\,$5$) templates, all of which are taken from low-$z$ observations or constructed from stellar population synthesis models and then subsequently applied to the high-$z$ universe. This small amount of templates is mostly due to observational constraints, as obtaining high signal-to-noise (S/N) spectra with broad wavelength coverage and accurate flux calibration is intrinsically difficult, even at low redshift. However, it is also due to a tendency to introduce physically-valid solutions by only considering non-negative modifications to pre-existing templates (i.e. by only allowing for \textit{additional} contributions from, e.g., single stellar populations, dust attenuation, emission line fluxes). This ``additive approach'' to template modification tends to limit templates to include only ``baseline'' galaxies that have intrinsically little reddening and weak emission lines (or where these features have been removed through some SED fitting process) in order to avoid missing regions of parameter space when constructing the model photometric grid.

Based in part on these constraints, traditional template-based approaches tend to be highly model-dependent, relying on simplistic modeling assumptions to extend (small) well-understood, well-sampled regions of parameter space into (large) less-understood, less-sampled ones, often generating on the order of $\sim$\,40 modified rest-frame templates for every baseline template. This is especially true when attempting to model many of the highly-extincted, intensely star-forming, and strongly emitting galaxies seen at higher redshifts \citep[e.g., ][]{masters+14,atek+14}, which often requires large modifications to the pre-existing template set based on simplistic modeling assumptions and locally-calibrated scaling relations \citep{ilbert+09}.

These caveats aside, however, traditional template-fitting approaches have been successfully used in a wide range of surveys such as, e.g., the Cosmological Origins Survey \citep[COSMOS;][]{scoville+07,capak+07,ilbert+09}, the Cosmic Assembly Near-infrared Deep Extragalactic Legacy Survey \citep{grogin+11,koekemoer+11,dahlen+13}, and the NEWFIRM Medium Band Survey \citep[NMBS;][]{whitaker+11,whitaker+12}, among others.

\subsection{The ``Archetype'' Approach}
\label{subsec:phot_archetype}

In the traditional template-fitting approach, we use a small set of local, well-understood templates and extend them to alternate regions of color space in a strongly model-dependent way in an attempt to \textit{simulate} the color-space manifold where galaxies live at higher redshifts. However, for more specific applications where more templates are available (e.g., searches at $z < 1$ over a more restricted wavelength range), we can instead attempt to sample the entirety of color space directly using a much larger discrete set of galaxy spectral ``archetypes'' without allowing for any variation. While the physical properties of these archetypes might not be as well ``understood'' as templates we can construct from scratch, this does not prevent us from using them to derive photo-z estimates.

One of the main benefits of using archetypes is that they are actual \textit{physically-realized} SEDs with complex dust physics, interstellar medium (ISM) emission, fully populated stellar populations, and real star formation histories. Given a large enough sampling of color space using \textit{many} archetypes, we can then construct a brute-force sampling of color space spanned by \textit{real} galaxies without needing to recourse to linear combinations of simplistic, model-dependent atomic components.

By relying on archetypes, we are able to use well-motivated modeling assumptions concerning the ``smoothness'' of the photo-z problem (in terms of moving templates to different redshifts) without being strongly dependent on nuisance parameters \citep[such as the effects of dust and emission lines;][]{ilbert+13,arnouts+13,wilkinson+15,momcheva+15} that plague traditional template-fitting approaches. The larger number of templates also allows the use of adaptive model weights based on population statistics, where the density of archetypes in color space as well as individualized priors can be used during the fitting process to modify the final $P(z)$ distribution, as in \citet{cool+13}.

The largest application of this approach was in the PRIsm MUlti-object Survey \citep[PRIMUS; ][]{coil+11,cool+13}, where the higher spectral resolution of the PRIMUS dataset (relative to most photometric surveys) necessitated the use of a larger set of archetypes in order to model observed features. Using a large sample of spectral templates from the Active Galactic Nuclei (AGN) and Galaxy Evolution Survey \citep[AGES; ][]{kochanek+12}, \citet{cool+13} took advantage the large number of available archetypes to derive both a reduced set of effective templates and a set of associated weights. These were then used to derive $P(z)$'s to their objects using a brute-force search over a two-tiered redshift grid (coarse then fine), achieving a typical point estimate precision of $\sigma_z/(1+z)<0.5\%$.

While generally effective, using archetypes this way have four main drawbacks:
\begin{enumerate}
	\item \textit{Heavily dependent on flux calibration}. Archetypes must have good flux-calibrated spectra over the relevant wavelength range of interest that span the properties of the observed galaxy population to avoid introducing strong biases into the SED fitting process. Although PRIMUS was able to achieve both of these requirements due to its limited spectral and redshift coverage, this is much more difficult for larger photometric surveys where available photometry can sometimes span from the far UV to the far IR (e.g., COSMOS).
	\item \textit{Dense color space sampling required}. A limited set of archetypes imposes a granularity into the continuous set of SEDs. In principle, we need to sample color space with enough archetypes to achieve a density higher than the typical color errors to avoid introducing biases for specific galaxy types. This constitutes a major observational challenge, especially with respect to sampling minor variations in emission line strengths and dust screens.
	\item \textit{Obtaining spectra time-consuming and biased}. Although we can attempt to obtain a representative sample of the underlying galaxy population of interest, the set of archetypes does not account for possible deviations away from any specific template due to varying dust extinction and/or emission line strengths, which become more significant as we move towards higher redshifts where spectra become exceedingly sparse and biased. This makes ``filling in'' color space more difficult.
	\item \textit{Brute-force approaches computationally infeasible}. An archetype-dependent approach fundamentally relies on exploring a large number of possible spectra as a function of redshift. This imposes heavy restrictions on the use of brute-force approaches to deriving $P(z)$, often necessitating some type of template-reduction process that consequently re-introduces many of the issues outlined above \citep[see, e.g.,][]{cool+13}.
\end{enumerate}

While archetype approaches are noticeably less common in the literature, they nonetheless represent an efficient way of deriving $P(z)$'s in a model-based but \textit{empirically-driven} manner, provided a suitable set of archetypes exist.

\subsection{The ``Fuzzy Archetype'' Approach}
\label{subsec:phot_fuzzy}

To bypass this granularity problem, we introduce a hybrid approach that takes advantage of the much broader range of parameter space afforded by the use of large numbers of archetypes while still allowing for limited but useful model-dependent results due to variation in dust and emission line strengths that can effectively ``fill in'' the gaps in color space coverage. We accomplish this by promoting each baseline archetype to a fuzzy ``cloud'' in color space using small perturbations in dust extinction and emission line strengths. By keeping the perturbations small, we limit the ability of the model to cross into physically unrealized SEDs. At the same time, the individual amount of fuzz for each archetype can be tuned based on the sampling density in a given region of color space.

To create these new ``fuzzy archetypes'', we break the original set of $\rom{N}{gal}$ archetypes into a set of baseline templates and perturbative emission line templates. After superimposing a set of individualized priors on the allowed variation in equivalent width (EW; i.e. continuum-normalized emission-line strength) $P(\Delta \rom{\textrm{EW}}{line})$ as well as dust attenuation $P(\Delta \rom{\ebv}{dust})$ for a given dust curve for each individual galaxy, we allow each small deviations away from each baseline archetype according to the underlying continuum shapes and emission line strengths.

This simple process transforms an individual galaxy's position in color space into a multidimensional PDF centered at its original location, enabling a given template set to more accurately capture the observed variability between galaxies of similar types. By allowing for small deviations around a set of well-chosen empirical templates that directly probe the regions of interest, fuzzy archetypes remain in the regime where assumptions such as a uniform dust screen approximation are likely to remain valid, helping to ensure that photometry corresponds with \textit{physically occupied} regions of color space. In addition, by placing priors on perturbations around observed emission line strengths, fuzzy archetypes move away from average scaling relations tied to multiple lines and/or other physical properties \citep[see, e.g., ][]{ilbert+09,salmon+15}, allowing them to probe greater (and more physical) variability in emission line strengths independently of one another. 

To take full advantage of this approach, we eschew using the limited collection of templates from \citet{coleman+80}, \citet{kinney+96}, or \citet{polletta+07} used in traditional template-fitting approaches (as well as the stellar population synthesis models often used to fill in missing color-space coverage). Instead, we utilize the set of 129 high-quality UV\,--\,IR spectra taken from \citet{brown+14}. The extensive multi-wavelength range spanned by their large collection of galaxy templates is ideal for our purposes, probing a wide range of color space, galaxy types, and emission line strengths. These features are briefly described and illustrated in Appendix~\ref{ap:brown} (Figure~\ref{fig:brown_templates} and Table~\ref{tab:brown_info}); we direct the reader there as well as to the original \citet{brown+14} paper for additional details about the template set.

\begin{figure}
	\centering
	\includegraphics[width=\linewidth]{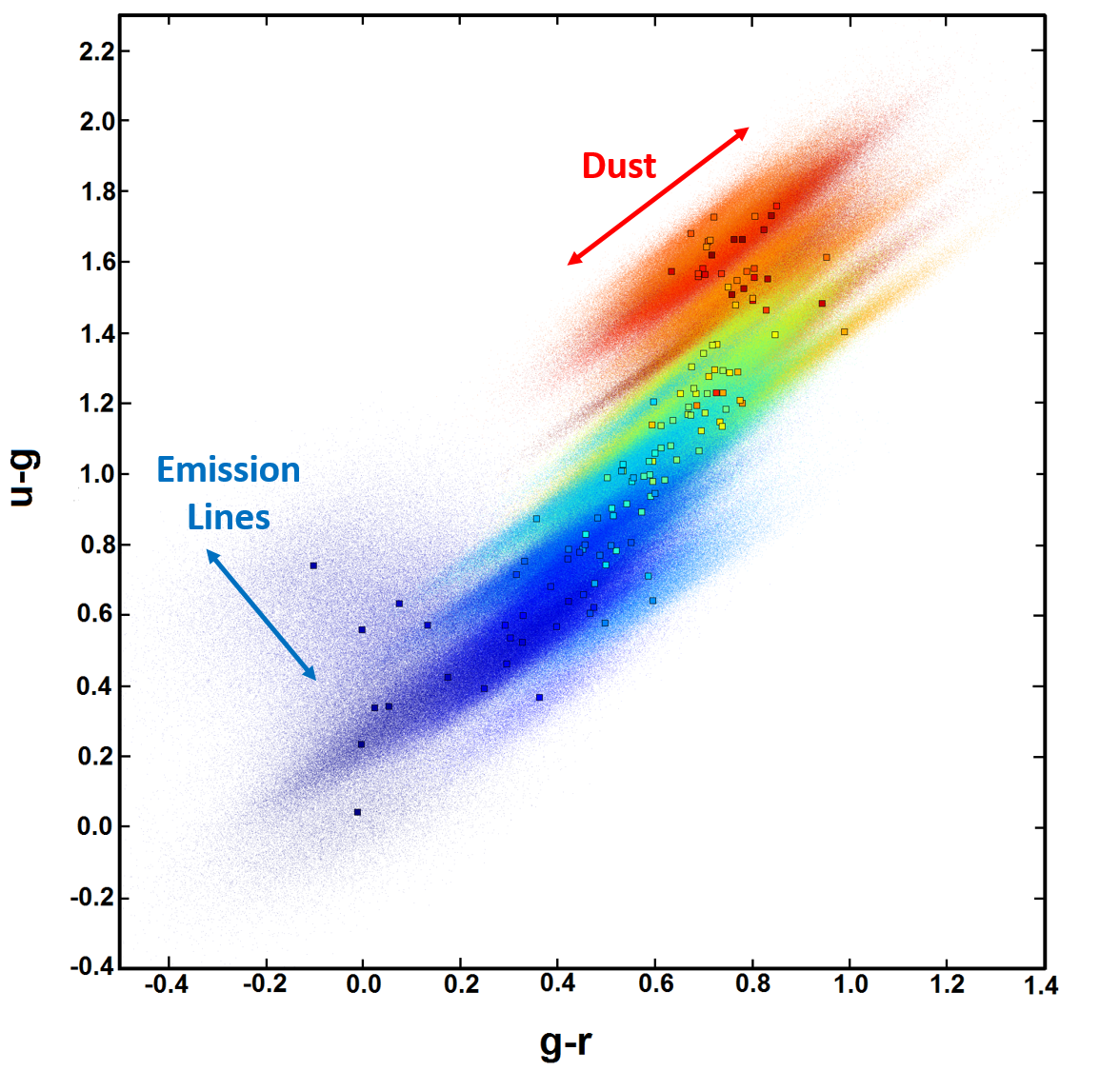}
	\caption{
		The rest-frame $u-g$ versus $g-r$ distribution (in magnitudes) for the 129 \citet{brown+14} templates, colored in order of increasing FUV emission (see Appendix~\ref{ap:brown}). For each template we have drawn 25,000 points (5000 for each dust template; see \S\ref{subsec:phot_fuzzy}) from the associated set of Gaussian priors $P(\Delta \ebv)=\mathcal{N}[\mu=0,\sigma=0.1]$ and $P(\Delta \textrm{EW})=\mathcal{N}[\mu=0,\sigma=0.2 \times \textrm{max}(\textrm{EW},0.005\rom{\lambda}{line})]$ corresponding to a toy set of ``fuzzy archetypes''. The impact of our priors is labeled, and can additionally be seen by examining the distribution of the redder (weakly emitting) galaxies as compare to the bluer (strongly emitting) galaxies. While the extent of reddening allowed by our priors dominates the majority of color variation for redder galaxies, emission line variation has a comparable effect in bluer ones. This illustrates the significant impact emission lines can have on the observed SED in highly active star-forming galaxies.}
	\label{fig:fuzzy_templates_ug_gr}
\end{figure}

For each galaxy, we measure the corresponding EWs of the $\lbrace$H$\alpha$+N{\scriptsize[II]} complex, H$\beta$, H$\gamma$, O{\scriptsize[II]}, O{\scriptsize[III]}$_{5008}$, O{\scriptsize[III]}$_{4960}\rbrace$, with corresponding effective central wavelengths of $\rom{\lambda}{line} = $ $\lbrace 6564.6$, $4862.7$, $4341.7$, $3727.0$, $5008.2$, $4960.3\rbrace$ and an assigned $\Delta v = \pm$ $\lbrace 2000$, $1500$, $1500$, $1500$, $1500$, $1250\rbrace$\,km/s. We model the continuum using a linear fit over the regions contained within $\pm 100$\,--\,$155\%$ of the corresponding $\Delta v$ spread. After combining the two O{\scriptsize[III]} lines into a single template, we are left with five emission line templates per archetype. Our derived EWs for each galaxy, ordered in terms of increasing FUV flux, are listed in Table~\ref{tab:brown_info}.

\begin{figure*}
	\centering
	\includegraphics[width=\linewidth]{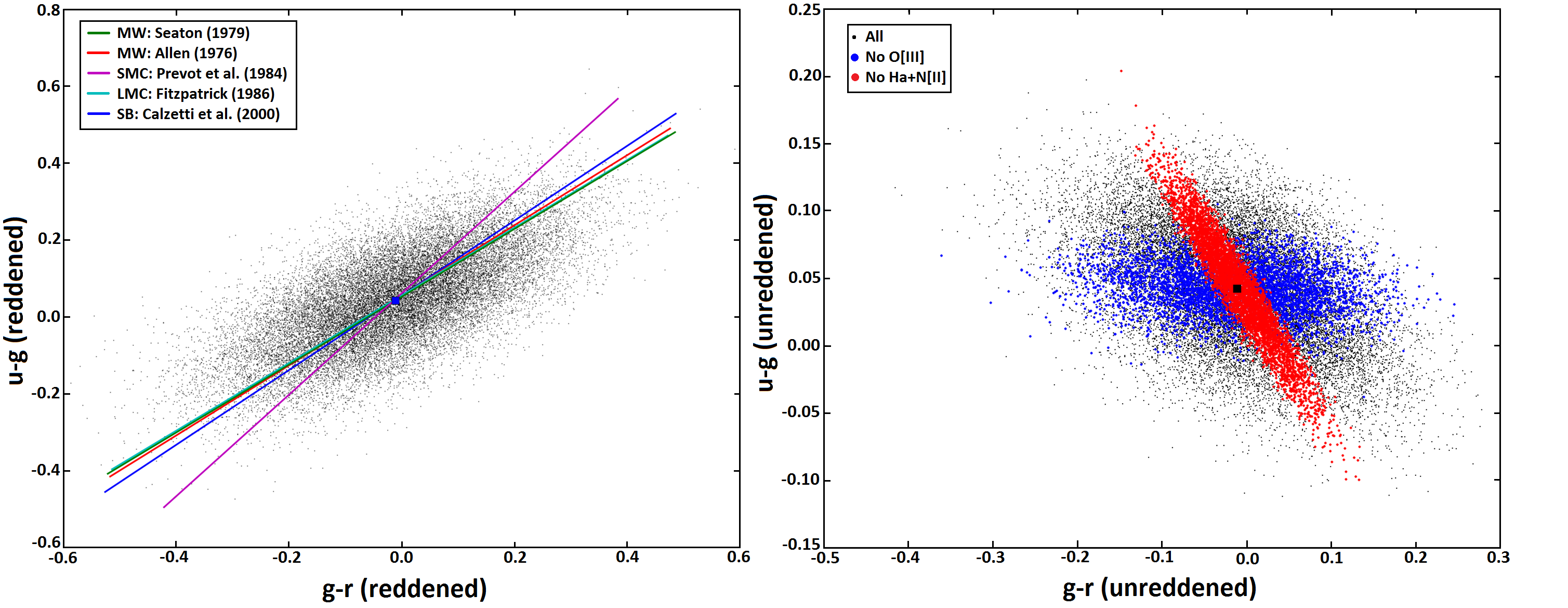}
	\caption{
		\textbf{Left panel:} The $u-g$ versus $g-r$ distribution (in magnitudes) for UGCA 166, the strongest H$\alpha$+N{\scriptsize[II]} emitter in our sample with a corresponding equivalent width of $\sim$\,860\,{\AA}. The ``reddening vectors'' of each of the five standard dust curves employed in the literature (see \S\ref{subsec:phot_fuzzy}) in the $u-g$ versus $g-r$ plane are overplotted as solid lines. The similarity between the majority of the dust curves is concerning and suggests the limited set of empirical dust templates in use today are not effective at probing the full available range of possible reddening curves.
		\textbf{Right panel:} The un-reddened $u-g$ versus $g-r$ distribution. Subsamples excluding $>5\%$ variation in O{\scriptsize[III]} (blue) and the H$\alpha$+N{\scriptsize[II]} complex (red) demonstrate that almost all of the ``fuzziness'' in UGCA 166 is driven by variations in H$\alpha$+N{\scriptsize[II]} (via the $r$ band; horizontal) and O{\scriptsize[III]} (via the $g$ band; diagonal), with lesser contributions from H$\beta$, H$\gamma$, and O{\scriptsize[II]}, further illustrating the large impact emission lines can have on observed colors.}
	\label{fig:ugca_166_fuzzy}
\end{figure*}

To demonstrate the ``blurring'' effect that our fuzzy archetypes have on the relevant regions of color space, we illustrate the extent to which a set of dust templates and an simplistic set of priors can affect the color distribution. Using the same set of five standard dust templates utilized in previous photo-z searches \citep{bolzonella+00,ilbert+09,speagle+14,speagle+15} -- two Milky Way (MW) templates \citep{allen76,seaton79}, a Large Magellanic Cloud (LMC) template \citep{fitzpatrick86}, a Small Magellanic Cloud (SMC) template \citep{prevot+84}, and starburst (SB) template \citep{calzetti+00} -- we assign the set of Gaussian priors $P(\Delta \ebv)=\mathcal{N}[0,0.1]$ and $P(\Delta \textrm{EW})=\mathcal{N}[0,0.2 \times \max(\textrm{EW},0.005\rom{\lambda}{line})]$ () to all archetypes and associated emission line templates, where the notation $\mathcal{N}[\mu,\sigma]$ signifies a Gaussian (i.e. Normal) distribution with mean $\mu$ and standard deviation $\sigma$. We choose these priors to illustrate the range in extinction probed by current codes (which often allow $\ebv \geq 0.5$) as well as intrinsic variability in emission line scaling relations (which are on the order of $\pm 0.2$\,dex; \citet{kennicutt98}). We then draw 5,000 samples from the priors for each archetype and dust template (a total of 25,000 samples per archetype), generate photometry using the convolution detailed in equation (\ref{eq:phot}), and plot the output $u-g$ and $g-r$ colors for the entire collection of samples in Figure~\ref{fig:fuzzy_templates_ug_gr}.

Although our choice of priors is overly simplistic, with the large changes in attenuation caused by our choice of $P(\ebv)$ leading to a thin, extended locus for red galaxies, the general blurring effect we wished to demonstrate is quite clear. This is especially true for bluer galaxies where the larger emission line contributions lead to comparable variations in color.

We illustrate these effects for the strongest H$\alpha$+N{\scriptsize[II]} emitter in the sample (UGCA 166) in Figure~\ref{fig:ugca_166_fuzzy}. We find that, at least in our chosen set of colors, a majority of the dust templates lead to very similar slopes of the corresponding color-space ``reddening vector''. In addition, we observe a large gap between the SMC reddening vector compared to the rest of the chosen dust templates. This suggests that the standard set of dust templates used in the literature are not well suited to probing the full region of allowed parameter space that brackets the observed reddening vectors \citep[see also][]{boquien+09}. We return to this issue in \S\ref{subsec:final_fuzzy}.

We also can observe the impact variation in individual emission lines can have on the observed color of UGCA 166. In particular, variation in H$\alpha$+N{\scriptsize[II]} from our 20\% prior leads to changes in the observed color of up to $\sim 0.4$\,mag (via the $r$ band), while variation in O{\scriptsize[III]} leads to more modest changes of up to $\sim 0.1$\,mag (via the $g$ band). As different galaxies have a range of emission line strengths, the impact their variation can have on observed galaxy color as a function of redshift will vary widely.


\subsection{Linearized Fuzzy Templates}
\label{subsec:linear_fuzzy}

While the basic fuzzy template framework outlined above is straightforward, it remains difficult to incorporate into pre-existing fitting schemes because both probing and marginalizing over the fuzziness of the template must be done numerically via equation (\ref{eq:phot}). In essence, the only methods that could effectively utilize fuzzy archetypes as outlined above would rely on pre-generated photometry based on extensive MC sampling of the priors, turning an initial set of $N$ baseline archetypes into a much larger set of $M \gg N$ discrete samples from the fuzzy archetypes. This precludes implementations that might involve extensive MC sampling as a function of redshift, can quickly marginalize over nuisance parameters and priors, and/or can take advantage of linear fitting schemes.

To allow for more flexible fitting approaches, we can approximate the perturbative impact of dust and emission lines using a first-order expansion around our baseline archetypes. This allows us to decompose our original model photometry at a given redshift $\rom{\mathbf{F}}{model}(z)$ into a linear combination of galaxy, emission line, and dust components. These components can then be pre-generated on a given redshift grid and simply read into memory, where they can be quickly combined to generate new model photometry (and thus large mock catalogs).

This process also turns SED fitting with fuzzy templates into a linear algebra problem that involves solving a system of linear equations with associated weights and priors. The best-fit answer can be quickly computed using simple modifications/extensions to linear least squares fitting techniques, giving us a way to quickly marginalize over our linear parameters (the impact of dust and emission line variation) for a given archetype and dust template at a particular redshift. 

In practice, we feel that meeting this linearization requirement is a good guideline to effectively utilizing fuzzy templates -- in order to solve the issues highlighted in \S\ref{subsec:phot_archetype} and \S\ref{subsec:phot_fuzzy}, there should enough archetypes to enable tight priors such that linear approximations to their impact on the SED remain valid. We thus briefly outline the general fitting methodology below for completeness. In this paper, however, we actually focus on large-scale MC sampling of the fuzz discussed above because we are able to sort them in a manner that speeds up the complex search in SED-redshift space.

\subsubsection{Linear Approximation}
\label{subsubsec:linear_fuzzy_1}
Ignoring rest-frame dust attenuation and linear combinations of templates (see \S\ref{subsec:phot_archetype}), the ``baseline'' model photometry $\rom{\mathbf{F}}{base,gal}$ for a galaxy at a given redshift is
\begin{equation}
s\rom{\mathbf{F}}{base,gal}(z)=s\rom{\mathbf{F}}{gal}(z) + \sum_{\textrm{lines}} \Delta \textrm{EW}_{\textrm{lines}}^\textrm{gal} \,s\rom{\mathbf{F}}{lines}^\textrm{gal}(z),
\end{equation}
where $\rom{\mathbf{F}}{gal}(z)$ and $\rom{\mathbf{F}}{lines}^\textrm{gal}(z)$ are the photometric fluxes (\textit{including dust attenuation from the IGM}) for each galaxy and its corresponding set of emission line templates and we have included the general scaling factor $s$ to emphasize that our photometry must be properly scaled before comparing with observed photometry $\rom{\mathbf{F}}{obs}$.

Although the impact of attenuation from rest-frame dust screen for a given template is wavelength-dependent, its impact in each band can be approximated as $\exp\left[{\rom{\mathbfcal{R}}{dust}(z)\Delta\ebv}\right]$, where $\rom{\mathbfcal{R}}{dust}(z)$ is a collection of negative numbers determined by the given dust template $\rom{k}{dust}(\nu)$ and the corresponding filter set $\mathbf{T}(\nu)$ derived at a given redshift. Expanding about $0$ and ignoring all $\mathcal{O}(> 1)$ terms, we get
\begin{eqnarray}
s\rom{\mathbf{F}}{model}(z)= s\rom{\mathbf{F}}{gal}(z) + \Delta\ebv \,s\rom{\mathbf{F}}{dust}(z) \nonumber \\
+  \sum_{\textrm{lines}} \Delta \textrm{EW}_{\textrm{lines}}^\textrm{gal}\,s\rom{\mathbf{F}}{lines}^\textrm{gal}(z),
\end{eqnarray}
where $\rom{\mathbf{F}}{dust}(z) = \rom{\mathbfcal{R}}{dust}(z)\odot\rom{\mathbf{F}}{gal}(z)$ is our new ``dust photometry'' term.

\subsubsection{Linear Fitting}
\label{subsubsec:linear_fuzzy_fit}


The corresponding log-likelihood when fitting normally distributed data with a series of independent Gaussian priors is equivalent to a modified $\chi^2$ metric,
\begin{equation}\label{eq:chi2_mod}
\rom{\chi^2}{mod}(\boldsymbol{\theta},\boldsymbol{\phi},s) \equiv \sum_i \sigma_{i}^{-2} \left[\Delta F_i(\boldsymbol{\theta},\boldsymbol{\phi},s)\right]^2 + \sum_j \left(\frac{\phi_j}{\sigma_{\phi_j}(\boldsymbol{\theta})}\right)^2
\end{equation}
where $\sigma_i^2=\sigma_{\textrm{obs},i}^2+\sigma_{\textrm{model},i}^2$ is the $i$th component of the total variance,\footnote{For the remainder of the paper, we will assume $\rom{\boldsymbol{\sigma}}{model}=0$ such that $\boldsymbol{\sigma}=\rom{\boldsymbol{\sigma}}{obs}$.} $\Delta F_i(\boldsymbol{\theta},\boldsymbol{\phi},s)=F_{\textrm{obs},i}-sF_{\textrm{model},i}(\boldsymbol{\theta},\boldsymbol{\phi})$ is the $i$th flux residual, $\boldsymbol{\theta}=\left\lbrace z,\textrm{gal},\textrm{dust} \right\rbrace$ contains the non-linear parameters of interest, $\boldsymbol{\phi}=\left\lbrace \ebv, c_b^\prime, \lbrace\Delta \textrm{EW}\rbrace_{\textrm{lines}} \right\rbrace$ contains the linear nuisance parameters, $\sigma_{\phi_j}(\boldsymbol{\theta})$ is the standard deviation of the corresponding Gaussian prior for a given $\boldsymbol{\theta}$, the sum over $i$ is taken over all observed bands, and the sum over $j$ is taken over all relevant nuisance parameters.

\begin{figure}
	\centering
	\includegraphics[width=\linewidth]{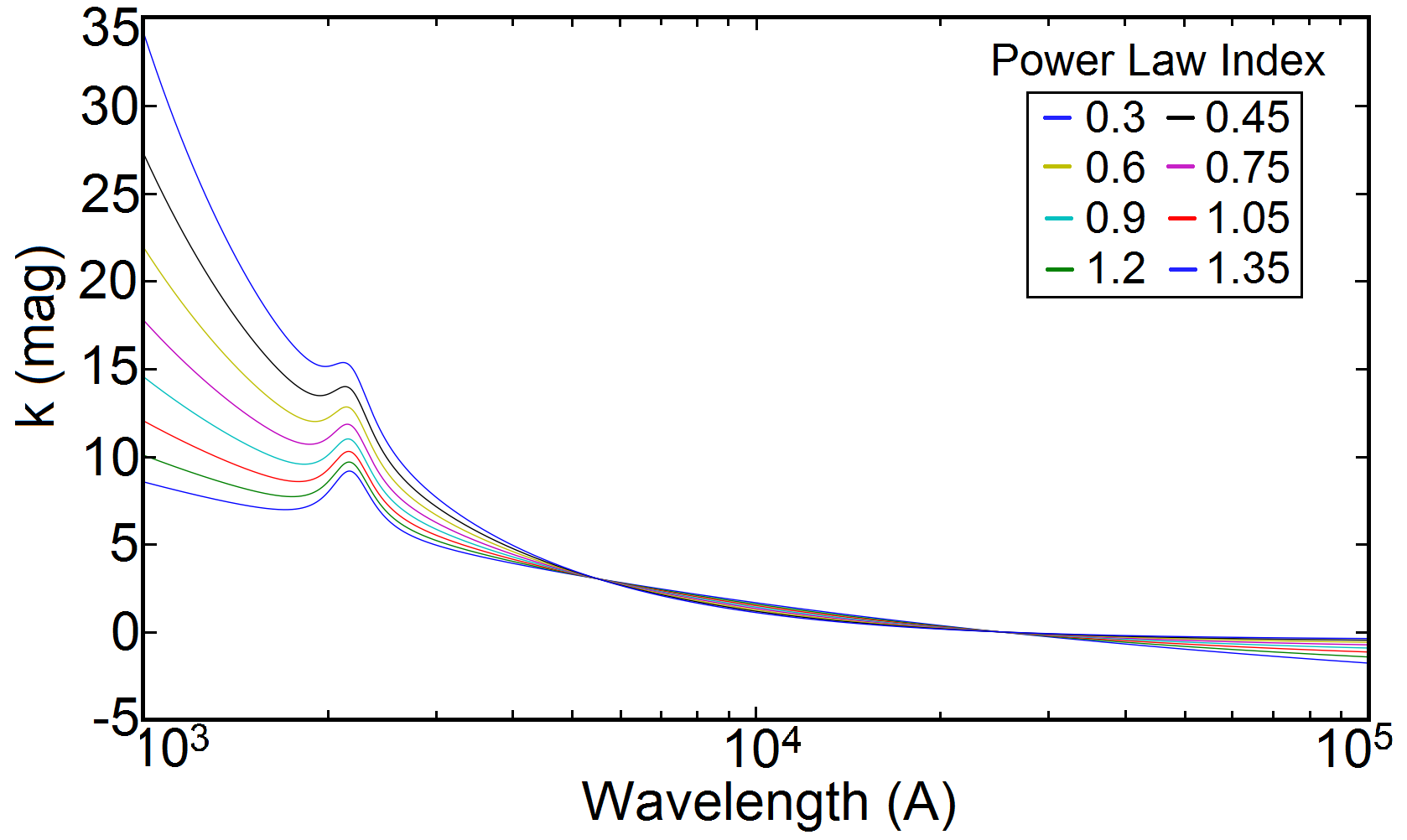}
	\caption{
		The eight dust attenuation curves used in this work. The underlying shape is a simple power law, $k(\lambda)=a\lambda^{-\beta}+b$, where $\beta$ brackets the general shapes of the starburst \citep{calzetti+00} and Small Magellanic Cloud \citep{prevot+84} attenuation curves and $k(\lambda)$ is normalized such that $A(V)=3.0$ (scaled) and $A(2.5\,\textrm{\um})=0$ (offset). The 2175\,{\AA} bump is superimposed on each of the underlying dust curves using the mean amplitude ($c_b$), position ($x_0$), and width ($\gamma$) of 328 MW curves taken from \citet{fitzpatrickmassa07}. The expanded range of parameter space that can be probed by these dust curves -- both with regards to changes to the underlying shape of the continuum as well as the overall contribution from the 2175\,{\AA} bump -- increases the power and flexibility of our fuzzy archetype-based approach.}
	\label{fig:powlaw_dust}
\end{figure}

For $\boldsymbol{\theta}$ and $\boldsymbol{\phi}$ fixed, we can marginalize over $s$ to minimize $\rom{\chi^2}{mod}(s|\boldsymbol{\theta},\boldsymbol{\phi})$, giving us
\begin{equation}\label{eq:chi_s}
s = \left. {\sum_i {\sigma_i^{-2}}} F_{\textrm{obs},i}F_{\textrm{model},i} \middle/ {\sum_i \sigma_i^{-2} F_{\textrm{model},i}F_{\textrm{model},i}} \right. ,
\end{equation}
which can be calculated prior to computing the actual $\rom{\chi^2}{mod}$ value.

While marginalizing over $s$ is a simple one-step process, marginalizing over the linear nuisance parameters $\boldsymbol{\phi}$ for fixed $s$ and $\boldsymbol{\theta}$ to minimize $\rom{\chi^2}{mod}(\boldsymbol{\phi}|\boldsymbol{\theta},s)$ involves solving a system of linear equations of the form
\begin{equation}\label{eq:chi_lparams}
\left(\mathbf{X}(\boldsymbol{\theta})^T\rom{\mathbf{W}}{obs}\right)\Delta\rom{\mathbf{F}}{gal}(\boldsymbol{\theta})=\left(\mathbf{X}(\boldsymbol{\theta})^{T}\rom{\mathbf{W}}{obs}\mathbf{X}(\boldsymbol{\theta})+\mathbf{W}_\phi(\boldsymbol{\theta})\right)\boldsymbol{\phi},
\end{equation}
where $\mathbf{X}(\boldsymbol{\theta})$ is the $\rom{N}{filt} \times N_{\boldsymbol{\phi}}$ matrix of pre-computed filter coefficients for a given $\boldsymbol{\theta}$,  $\Delta\rom{\mathbf{F}}{gal}(\boldsymbol{\theta}) = \mathbf{F}_{\textrm{obs}}-s\rom{\mathbf{F}}{gal}(\boldsymbol{\theta})$ is the baseline galaxy flux residual, $\rom{\mathbf{W}}{obs} = \textrm{diag}(\dots,\sigma_i^{-2},\dots)$ is the associated observational weight matrix, $\mathbf{W}_\phi = \textrm{diag}(\dots,\sigma_{\phi_j}^{-2},\dots)$ is the prior weight matrix, and $T$ is the transpose operator. This can be solved using standard linear least squares fitting techniques.

Together, these forms enables us to minimize $\rom{\chi^2}{mod}(\boldsymbol{\theta},\boldsymbol{\phi},s)$ with respect to $s$ and $\boldsymbol{\phi}$ for a given choice of $\boldsymbol{\theta}$ through an iterative least-squares fitting procedure with updates to $s$ at each iteration. This can then be run until a suitable stopping criterion is reach (e.g., in $\Delta\chi^2$).


\subsection{Our Fiducial Fuzzy Template Set}
\label{subsec:final_fuzzy}

As mentioned in \S\ref{subsec:gen_phot} and shown in Figure~\ref{fig:ugca_166_fuzzy}, the very limited set of extragalactic dust templates used in the literature do not appear to span the available range of possible reddening vectors. As we are only probing perturbative effects of dust at the few-percent level, we opt to construct a broader range of dust templates that better mimics the observed spread in attenuation properties \citep{boquien+09,boquien+12} instead of relying on the five empirical templates described in \S\ref{subsec:phot_fuzzy}.

In particular, we consider a modified dust template of the form 
\begin{equation}
\rom{k}{dust}^\prime(x)=\rom{k}{dust}(x)+c_b\rom{k}{bump}(x|x_0,\gamma),
\end{equation}
where $x\equiv 1/\lambda$ is measured in \ium and $\rom{k}{bump}(x|x_0,\gamma)$ is a Lorentzian-like function taken from \citep{fitzpatrickmassa07} where $x_0$ and $\gamma$ the central position and width of the Lorentzian feature, respectively. This ``bump'', most often located around 2175\,{\AA}, is a common observed feature among both Galactic \citep{fitzpatrickmassa07} and extragalactic \citep{kriekconroy13,scoville+15} sources. As it's amplitude $c_b$ can vary widely, ignoring it's impact on observed photometry can significantly bias quantities derived through SED fitting \citep{kriekconroy13}.

We model the dust curve as a simple power law in $x$ such that $k(x)=a_x(\beta)(x)^{-\beta}$+$b_x(\beta)$, where $a_x$ and $b_x$ are chosen such that each dust curve is normalized (scaled) to $A(V)=3.0$, the mean value from the set of 328 MW curves observed in \citet{fitzpatrickmassa07}, and $A(2.5\,\textrm{\um})=0$ (zero-point). We generate eight dust curves with $\beta=\lbrace 0.3,0.45,0.6,0.75,0.9,1.05,1.2,1.35 \rbrace$ that brackets the range of observed dust properties. We then add in the 2175\,{\AA} bump with the corresponding mean amplitude, position, and width of the bump $\lbrace c_b,x_0,\gamma \rbrace=\lbrace 3.0,4.592\textrm{\,\ium},0.922\textrm{\,\ium}\rbrace$ from \citet{fitzpatrickmassa07} as the amplitude of the bump can be probed freely using a corresponding prior on $P(\Delta c_b)$. The corresponding series of dust curves are plotted in Figure~\ref{fig:powlaw_dust}.

Based on our investigations in previous sections, we decide to use the following priors to define a ``fiducial'' set of fuzzy archetypes for testing:
\begin{eqnarray}
P[\Delta\ebv]=\mathcal{N}[0,\Delta A(1500\textrm{\,{\AA}})=0.05\,\textrm{mag}], \nonumber \\
P[\Delta c_b]= \mathcal{N}[0,1.5], \\
P[\Delta \rom{\textrm{EW}}{line}]=\mathcal{N}[0,0.2 \times \max(\textrm{EW},0.005\rom{\lambda}{line})]. \nonumber
\end{eqnarray}

Our choice to fix $P[\Delta\ebv]$ to $A(1500\,{\AA})$ is to limit the amount of dust extinction in the UV to the approximately linear regime ($\sim$\,$5\%$ 1$\sigma$ variation) across the entire wavelength range while also accounting for the varying shapes of our dust templates. Likewise, our choice of prior for $\Delta c_b$ is to allow fluctuations to $0$ to take place at the 2$\sigma$ level, which enables us to fit dust curves without the impact of the 2175\,{\AA} bump while generally disfavoring possible non-physical solutions. Finally, we opt to leave $P(\Delta \rom{\textrm{EW}}{line})$ unchanged from before, which again allows for 20\% variation ($\sim 0.1$\,dex) in EW, mimicking observed scatter among emission line scaling relations \citep{kennicutt98,kennicuttevans12} while also establishing an EW ``floor'' that allows for small variation at the $\lesssim 1$\,--\,$2$\% level.

\section{Re-organizing Model Photometry Using Self-Organizing Maps}
\label{sec:som}

\subsection{Overview}
\label{subsec:som_overview}

The Self-Organizing Map \citep[SOM;][]{kohonen82,kohonen01} is an unsupervised machine-learning algorithm that projects high-dimensional data onto a lower-dimensional space using competitive training of a large set of ``cells'' in a way that preserves general topological features and correlations present in the higher-dimensional data. We summarize the main features of our specific implementation of the algorithm below, and direct the reader to \citet{carrascokindbrunner14} and \citet{masters+15} for more details concerning applications SOMs to photo-z's.

A SOM consists of a fixed number of cells $\rom{N}{cells}=\prod_i \rom{N}{cells,i}$, where the product over $i$ is taken over all available SOM dimensions, arranged on an arbitrary dimensional grid with $\rom{N}{cells,i}$ in each dimension. Each cell in the grid is assigned a position on the SOM $\rom{\mathbf{x}}{som}^{\textrm{cell}}$ and contains a cell model $\rom{\mathbf{F}}{som}^{\textrm{cell}}(t)$ of the same size as a vector from the training data (i.e. $\rom{N}{filter}$). Training then proceeds as follows:
\begin{enumerate}
	\item Initialize the cell models (most often randomly) and set the current iteration $t=0$.
	\item Draw (with replacement) a random object $\rom{\mathbf{F}}{obs}^{i}$ from the input dataset.
	\item Compute $\rom{s}{cell}(t)$ and $\chi^2\left(\rom{\mathbf{F}}{obs}^{i},\rom{\mathbf{F}}{som}^{\textrm{cell}}(t)|\rom{s}{cell}(t)\right)$ for every cell on the map.
	\item Select the best-matching cell $\rom{\mathbf{x}}{som}^{\textrm{best}}$ from the corresponding set of $\chi^2$'s.
	\item Update the cell models within the map according to an evolving \textit{learning rate} $\mathcal{A}(t)$ and \textit{neighborhood function} (i.e. kernel density) $\mathcal{H}(\rom{\mathbf{x}}{som}^{\textrm{best}},\rom{\mathbf{x}}{som}^{\textrm{cell}}|t)$ such that
	\begingroup\makeatletter\def\f@size{8.5}\check@mathfonts
	\begin{equation}
		\rom{\mathbf{F}}{som}^{\textrm{cell}}(t+1)=\rom{s}{cell}(t)\rom{\mathbf{F}}{som}^{\textrm{cell}}(t)+\mathcal{A}(t)\mathcal{H}(t)\left[\rom{\mathbf{F}}{obs}^{i}-\rom{s}{cell}(t)\rom{\mathbf{F}}{som}^{\textrm{cell}}\right].
	\end{equation}
	\endgroup
	\item Increment $t$ and repeat steps (ii) through (v) for a set number of training iterations $\rom{N}{iter}$.
\end{enumerate}
After training, objects are ``sorted'' onto the map by repeating steps (iii) and (iv) for every object in the input dataset and assigning them to the best-matching cell.


The SOM has a number of explicit free parameters that can be tuned during training. Although a wide range of parameterizations for $\mathcal{A}(t)$ and $\mathcal{H}(t)$ are possible, we utilize the following functional forms:
\begin{eqnarray}
\mathcal{A}(t)=a_0\left(\frac{a_1}{a_0}\right)^{t/\rom{N}{iter}}, \\
\mathcal{H}(\rom{\mathbf{x}}{som}^{\textrm{best}},\rom{\mathbf{x}}{som}^{\textrm{cell}}|t) = \exp\left(-\frac{||\rom{\mathbf{x}}{som}^{\textrm{best}}-\rom{\mathbf{x}}{som}^{\textrm{cell}}||^2}{\sigma^2(t)}\right), \textrm{ and} \\
\sigma(t)=\sigma_0\left(\frac{1}{\sigma_0}\right)^{t/\rom{N}{iter}}.
\end{eqnarray}
The \textit{hyper-parameters} $a_0$ and $a_1$ control the ``stiffness'' of the map when new data is added, with $a_1$ establishing the learning adjustment at the beginning of the training and $a_0$ the final threshold, while $\sigma_0$ controls the corresponding scale of the Gaussian kernel.

\subsection{Using SOMs on Model Spaces}
\label{subsec:som_model}

\begin{figure*}
	\centering
	\includegraphics[width=\linewidth]{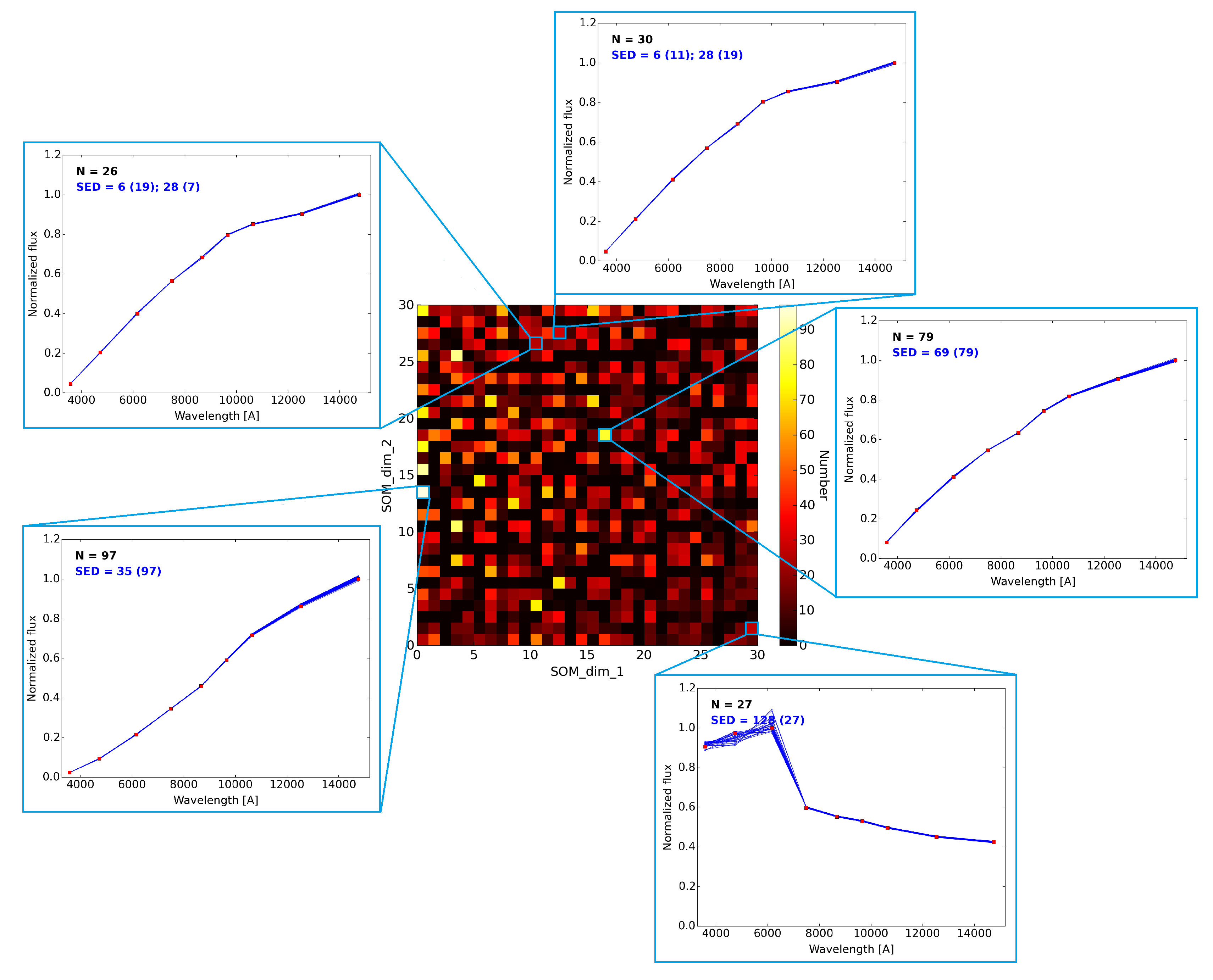}
	\caption{
		The $30 \times 30$ 2-D Self-Organizing Map (SOM) constructed from our rest-frame model catalog of 9,675 Monte Carlo realizations of our fuzzy archetypes uniformly sampled from our fiducial set of dust curves (Figure~\ref{fig:powlaw_dust}) and baseline archetypes (see \S\ref{subsec:final_fuzzy}), with each ``cell'' colored according to its occupation rate (i.e. the number of objects matched to a given cell). Individual insets around the map illustrate the corresponding cell model (red squares) as well as the model data contained within that cell (blue lines), with the number of objects and corresponding SED identifier (labeled in order of increasing normalized FUV flux) also shown along with the associated number of objects (listed in parenthesis). The SOM is able to effectively the dataset and preserve local topology such that nearby cells on the SOM and the objects contained within them are more similar to each other than those that are widely separated. The number density of objects sorted across the SOM provides information about the general distribution of our chosen set of 129 fuzzy \citet{brown+14} archetypes in color space, with higher number densities in the upper left region of the map illustrating the corresponding denser template coverage in color space seen in Figure~\ref{fig:brown_templates}.}
	\label{fig:2dsom_overview}
\end{figure*}

In most cases, SOMs are used on observed datasets to investigate their high-dimensional covariances and clustering properties in a way that's easy to visualize. However, we instead use the SOM as a way to better understand the \textit{model} space and its mappings to the corresponding set of observables. In doing so, we ``flip'' the typical view of template-fitting approaches from being entirely forward-mapping problems to partially inverse-mapping ones.

In essence, the naive approach to deriving the posterior distribution $P(\boldsymbol{\theta},\boldsymbol{\phi}|\rom{\mathbf{F}}{obs})$ for a given set of fuzzy archetypes relies on combining the set of forward mappings $\left\lbrace(\boldsymbol{\theta},\boldsymbol{\phi})\rightarrow\rom{\mathbf{F}}{model}\right\rbrace$ with the associated log-likelihoods $\left\lbrace\rom{\chi^2}{mod}(\rom{\mathbf{F}}{obs}|\rom{\mathbf{F}}{model}(\boldsymbol{\theta},\boldsymbol{\phi}),s)\right\rbrace$. Because computing $\rom{\mathbf{F}}{model}(\boldsymbol{\theta},\boldsymbol{\phi})$ from scratch is computationally expensive (\S\ref{subsec:gen_phot}) and the dependence of $\rom{\mathbf{F}}{model}$ on individual parameters is almost guaranteed to be highly skewed, covariant, and non-linear (\S\ref{subsec:phot_fuzzy}; although see \S\ref{subsec:linear_fuzzy}), forward-modeling approaches often must to resort to sophisticated sampling techniques to overcome these challenges \citep{foremanmackey+13,feroz+13,speagle+15}.

However, the \textit{inverse} problem of determining the likelihoods of possible predicted observables given the observed data (i.e. computing $\rom{\chi^2}{mod}(\rom{\mathbf{F}}{model}^\prime|\rom{\mathbf{F}}{obs})$ for a hypothetical set of observables $\rom{\mathbf{F}}{model}^\prime$) is easy since the set of associated uncertainties are known. If the corresponding inverse mapping(s) $\rom{\mathbf{F}}{model}^\prime\rightarrow(\boldsymbol{\theta},\boldsymbol{\phi})$ are also known, then exploring the set of possible $\rom{\mathbf{F}}{model}^\prime$'s allows rapid and \textit{exact} recovery of the posterior.

\begin{figure*}
	\centering
	\captionsetup[subfigure]{labelformat=empty}
	\subfloat[(a)][]{\includegraphics[width=\linewidth]{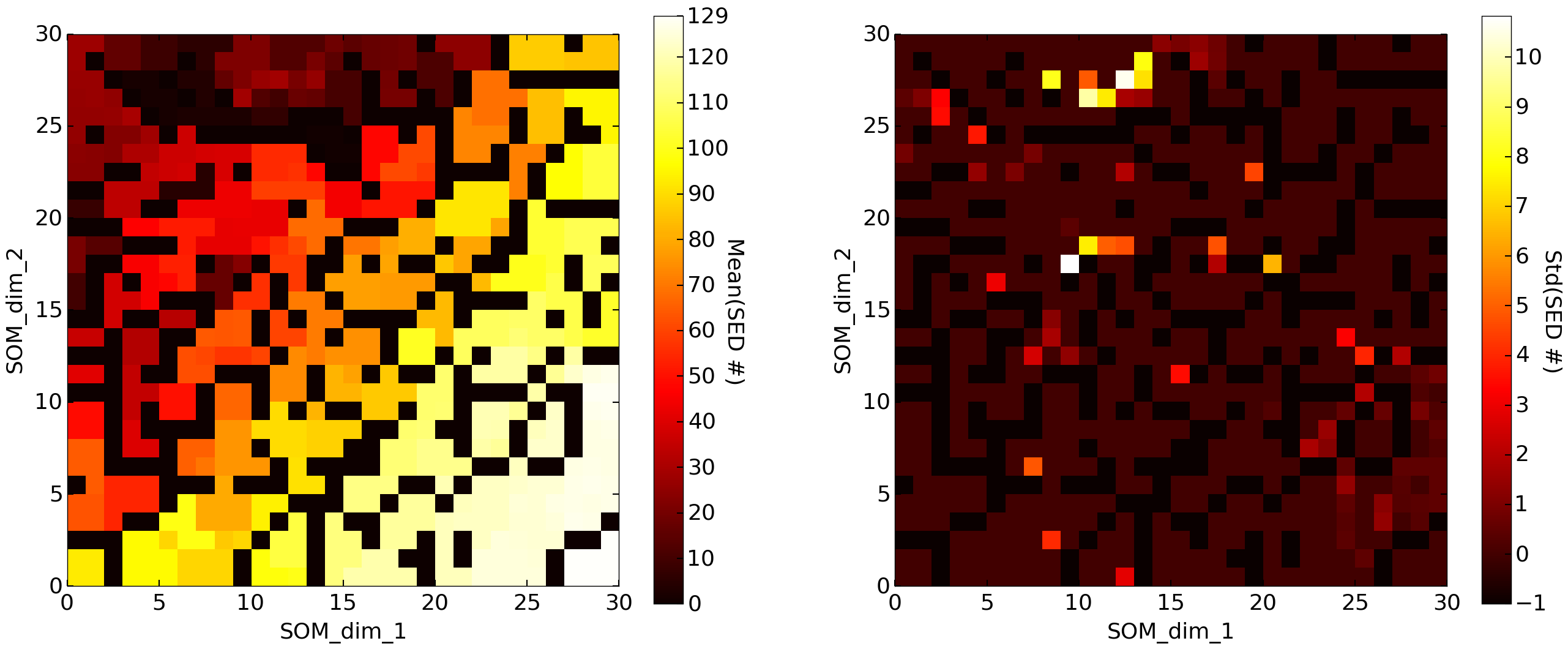}}
	\qquad
	\subfloat[(b)][]{\includegraphics[width=\linewidth]{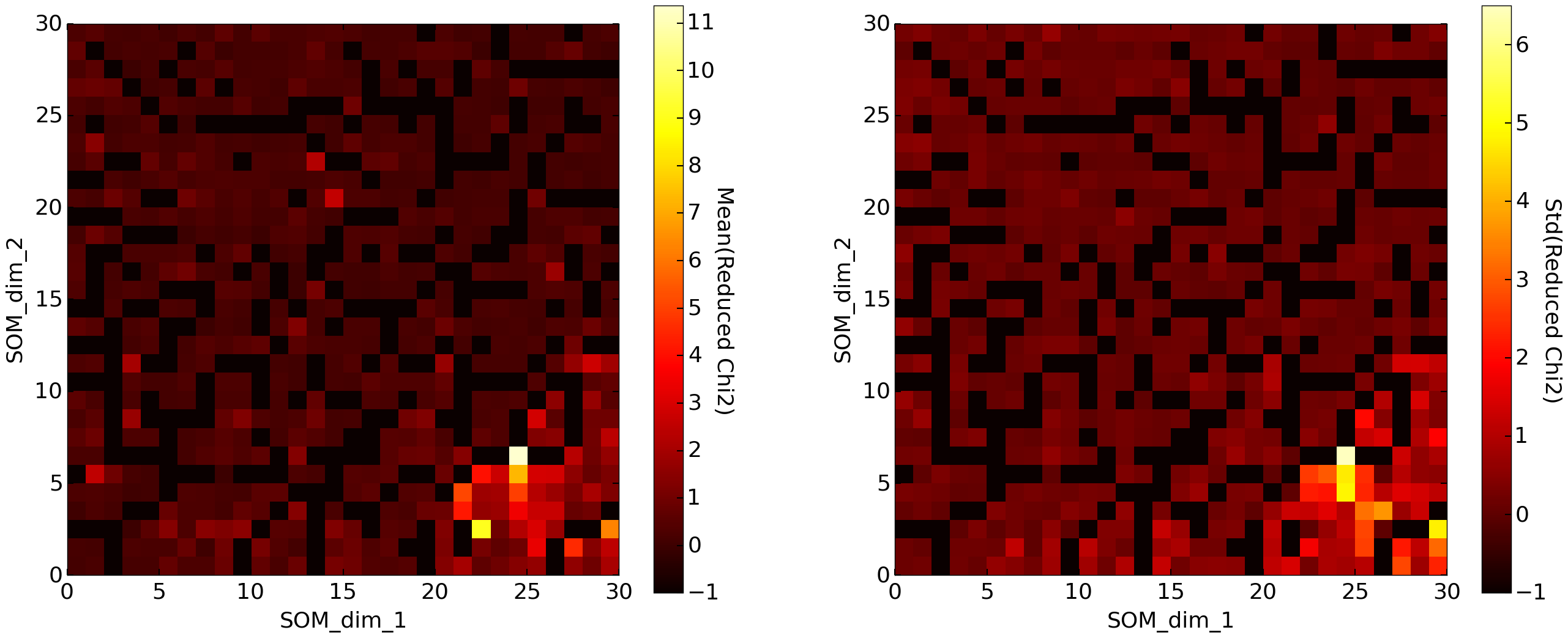}}
	\caption{
		\textbf{Top:} As Figure~\ref{fig:2dsom_overview}, but now colored according to the mean (left) and standard deviation (right) of the intrinsic SED distribution in a given cell. Cells with no data are assigned a value of $-1$. While our original 1-D SED ordering based on normalized FUV flux roughly captures the general behavior of the template set, the SOM is able reorganize the input parameter space into a coherent, smooth 2-D projection (left) where redder galaxies live in the upper left and bluer ones in the lower right. In addition, it is able to accurately capture ``wormholes'' -- regions of parameter space where objects that are (widely) separated in terms of input model parameters give rise to very similar colors -- in our rest-frame fuzzy templates (right), which appear as cells with high standard deviations.
		\textbf{Bottom:} As the top panel, but for reduced-$\chi^2$. As strong emission line variation produces large changes in color, the color variability of bluer galaxies in our input model is \textit{intrinsically larger} than redder galaxies. As the corresponding templates are also less common among our sample of archetypes (see Figure~\ref{fig:2dsom_overview}), the associated region is also subsequently more sparsely sampled by the SOM. This leads to an increase in the ``color volume'' spanned by these objects in a given SOM cell, causing the corresponding regions of the map to display both large mean $\chi^2$ values (left) and large standard deviations (right). For instance, the cell displaying the largest values in both panels almost exclusively contains realizations of the extreme emission line galaxy UGCA 166, also shown in Figure~\ref{fig:ugca_166_fuzzy}.}
	\label{fig:2d_som_2}
\end{figure*}

As a result, using the SOM as a form of \textit{discrete, non-linear dimensionality reduction} allows us to reconstruct the projection of inverse mappings generated from large ``model catalogs'' -- created from Monte Carlo sampling from our fuzzy archetypes and associated prior(s) $P(\boldsymbol{\theta})$ on our discrete/non-linear parameters -- that are clustered based upon their relevant \textit{predicted observables} rather than generative parameters. Once the appropriate SOM has been constructed and objects have been subsequently sorted onto it, we can save the entire collection of inverse relations and simply run any desired set of sampling techniques ``on top of'' the $N$-dimensional SOM space rather than the $\rom{N}{filt}$-dimensional input parameter space. This allows us to keep the \textit{entire} fitting process in (observed) color space, which we can explore using the lower-dimensional projection afforded by the SOM.

As we are interested in applications of this general methodology to future large-scale surveys, we choose our observables to have the same properties as the combined LSST and \textit{Euclid} wide-field surveys. This includes data in the LSST $ugrizY$ bands and the \textit{Euclid} $YJH$ bands (spanning $\sim$\,3500 to 15000\,{\AA}) with associated $5$-$\sigma$ imaging depths of 26.5\,mag (24.5 in $u$) for LSST and 24\,mag for \textit{Euclid} and an assumed calibration uncertainty of 0.01\,mag ($\sim$\,1\%).

\subsection{Application to Rest-Frame Photometry}
\label{subsec:som_rest}

We first investigate using the SOM to explore galaxy rest-frame colors in a more flexible manner. Pre-generated grids for model photometry often combine simple 1-D orderings for galaxy and dust templates in addition to a separate dimensions for the impact of $\ebv$ and emission line strengths to construct the multidimensional grid. This is the equivalent of combining a number of crude 1-D projections in order to probe a higher-dimensional space. While this is fine for brute-force approaches that fit every grid point, it can heavily impact the multidimensional gradients adaptive sampling techniques often rely on, leading to relatively ``bumpy'' photo-z likelihood surfaces \citep{speagle+15}. Allowing our fuzzy archetypes to live in intrinsically higher-dimensional manifolds established by the SOM clustering over rest-frame colors, however, should result in a much smoother likelihood surface -- especially when probed as a function of redshift.

The model catalog for we use for training is generated from the fiducial set of fuzzy archetypes established in \S\ref{subsec:final_fuzzy}.\footnote{All model catalogs explored in this paper are generated using the linear approximation scheme from \S\ref{subsec:linear_fuzzy} for computational convenience. We confirm the main results and conclusions remain unchanged if the full non-linear framework is used.} We assign each template a number based on the FUV flux (see Table~\ref{tab:brown_info}) of its corresponding archetype from 1 to 129, which we use to establish a crude 1-D ordering that we will use for comparison. For each of the 129 newly ordered fuzzy archetypes, we generate 75 Monte Carlo realizations from $P(\phi)$ and uniform sampling of $\rom{k}{dust}$. We then assign $1$\% errors to each of the 9,675 output model photometric fluxes in order to ensure the SOM training proceeds in logarithmic rather than linear space due to the astrophysical nature of the problem.

We construct a two-dimensional $30 \times 30$ SOM on this model catalog, with the cell models initialized by uniform random sampling on the interval $[0,1]$. After experimenting with a range of options, we settle on $[a_0,a_1,\sigma_0]=[1,0.5,30]$ for our hyper-parameters (similar to \citealt{masters+15}) and allow the map to train for $\rom{N}{iter}=10,000$ iterations. After training, we sort the 9,675 objects in our model catalog back onto the SOM using the corresponding set of best-matching cells as determined by their $\chi^2$ values. The final 2-D rest-frame SOM, colored according to its \textit{occupation rate} $N(\rom{\mathbf{x}}{som}^{\textrm{cell}})$ (i.e. the number of objects matched to a given cell) and with additional insets showcasing individual cell models, is shown in Figure~\ref{fig:2dsom_overview}.

As evidenced by the individual cell models, the SOM does an efficient job of clustering the dataset, and is well suited to preserving local topology such that nearby cell models look more similar than those that are widely separated. In addition, the general distribution of objects on the SOM gives us information on the general distribution of our chosen set of fuzzy archetypes in color space. While the number density across the map in general appears fairly uniform -- as desired, since regions more densely occupied within the dataset are more heavily sampled during the training process -- higher average number densities can be found in the upper left region of the map, indicating the denser coverage of the \citet{brown+14} templates in certain regions of color space (see also Appendix~\ref{ap:brown}).

To more clearly illustrate this point, in Figure~\ref{fig:2d_som_2} we plot the SOM colored according to the mean and standard deviation of the $SED(\rom{\mathbf{x}}{som}^{\textrm{cell}})$ and reduced-$\chi^2(\rom{\mathbf{x}}{som}^{\textrm{cell}})$ distributions in each cell (top and bottom panels, respectively). As can be seen in the top left panel, the SOM reorganizes the input parameter space into a coherent, smooth 2-D projection. In particular, while our 1-D ordering scheme seems to roughly capture the general behavior of the template set (top left panel), we observe that the SOM is able to accurately capture (widely-separated) degeneracies in our fuzzy template space (top right panel), at least according to our assigned 1-D ordering, that we might otherwise have missed. Capturing these ``wormholes'' in parameter space where objects that are widely separated in terms of input model parameters give rise to very similar colors is crucial to properly constructing multi-modal PDFs. This becomes especially true when redshift is included, as explored in \S\ref{subsec:som_model}.

We can also observe the impact our priors on emission line variation have on the rest-frame observables using the $\chi^2(\rom{\mathbf{x}}{som}^{\textrm{cell}})$ distribution shown in the bottom two panels. As strong emission line variation produces large changes in color, the variability of bluer galaxies from our input priors is higher that redder ones. As the corresponding templates are also less common among our sample of archetypes (see Figures~\ref{fig:brown_templates} and~\ref{fig:2dsom_overview}), the corresponding region is subsequently more sparsely sampled by the SOM, leading to regions of the map with larger associated ``color volumes'' and thus higher errors introduced by the relevant discrete cell models (bottom left) as well as intrinsic variability (bottom right). This is not an indication of ineffective training, but rather \textit{real variation in color as a function of a galaxy's position in color space}.

In summary, we find that the SOM is quite effect at reorganizing the input rest-frame regions of color space occupied by our collection of fuzzy archetypes. In particular, it is able to capture intrinsic color variability, provide information about our distribution of templates, and find wormholes among crude 1-D template orderings. A 2-D SOM thus serves as an effective tool for a 2+1-D photo-z algorithm run over the 2-D SOM space with redshift included as a free parameter.



\begin{figure*}
	\centering
	\captionsetup[subfigure]{labelformat=empty}
	\subfloat[][]{\includegraphics[scale=0.20]{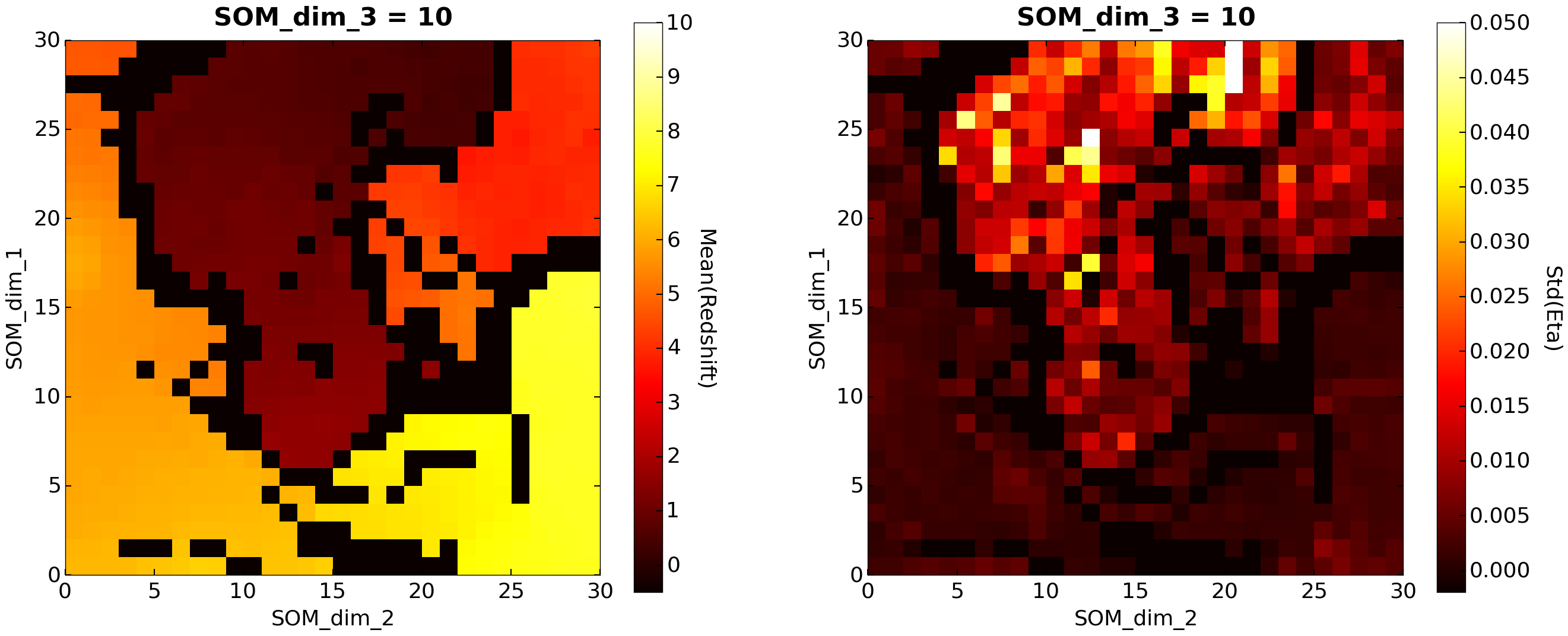}}
	\qquad
	\subfloat[][]{\includegraphics[scale=0.20]{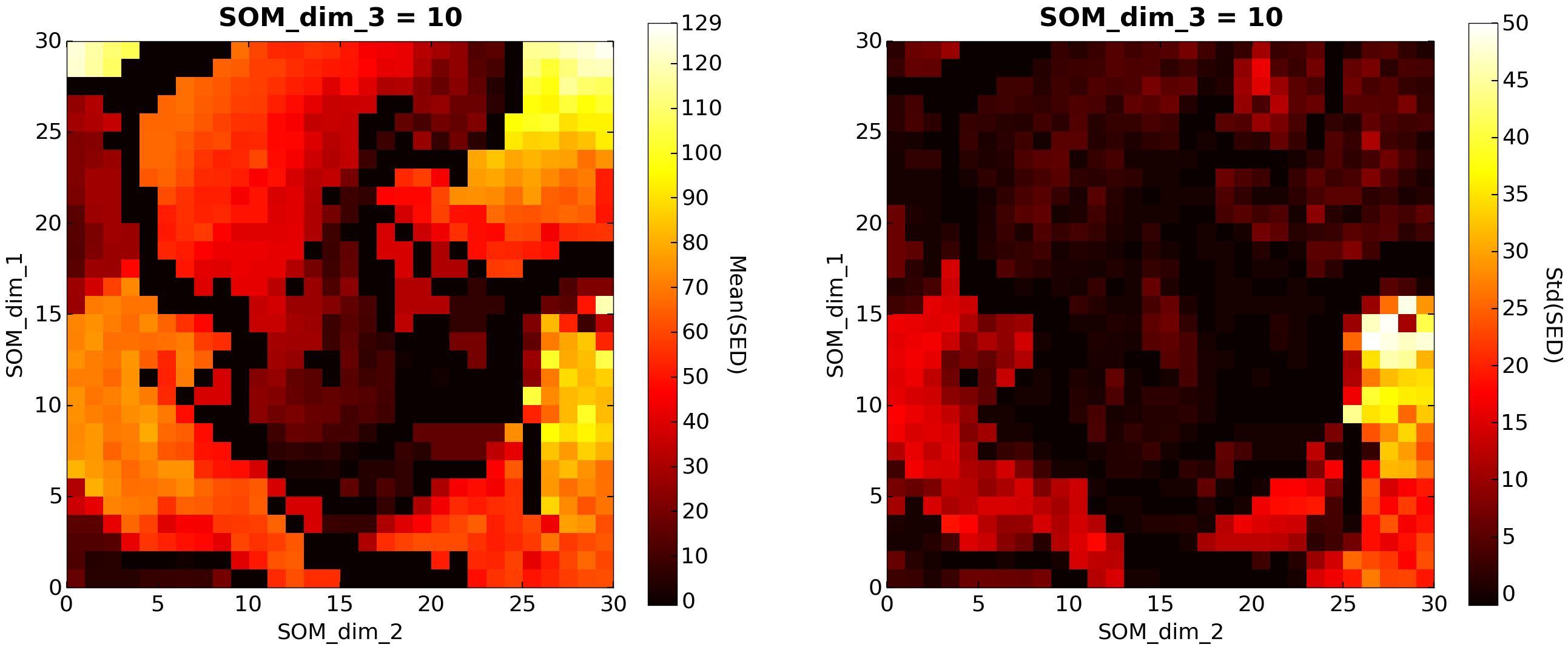}}
	\qquad
	\subfloat[][]{\includegraphics[scale=0.20]{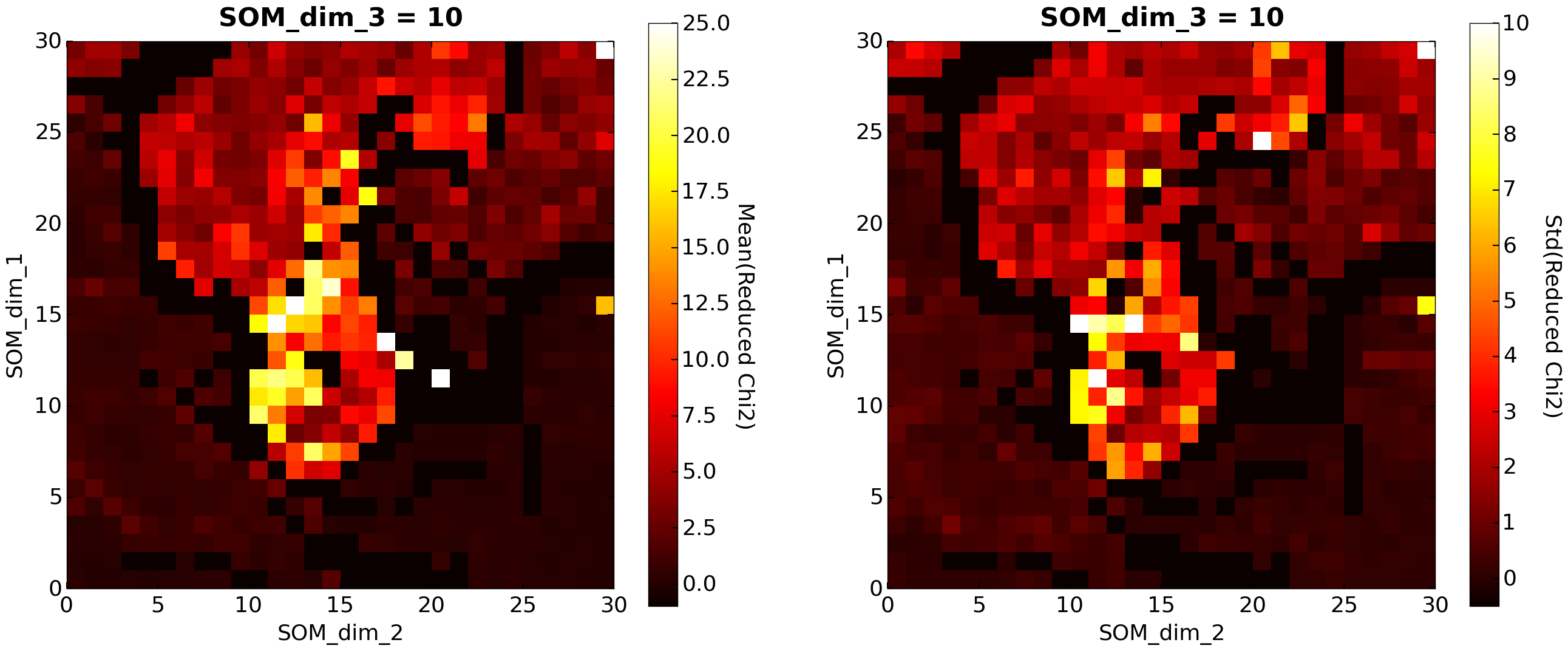}}
	\caption{
		As Figure~\ref{fig:2d_som_2}, but now illustrating a 2-D slice from the 3-D $30 \times 30 \times 30$ SOM constructed from our observed-frame model catalog of 3,228,225 Monte Carlo realizations of our fuzzy archetypes sampled from  $z=0$\,--\,$10$ ($\Delta z = 0.01$) and seeded with an appropriate set of $u-r$, $r-z$, and $z-H$ color gradients. 
		\textbf{Top:} Mean (left) and normalized standard deviation ($\eta\equiv\Delta z/(1+\bar{z})=0.05$; right) of the redshift distribution in a given cell. A smooth gradient in redshift evolution in different regions can be seen (left) along with the varying amount of information content contained in the color-redshift relation as a function of redshift (right). An upper limit of $\eta=0.05$ has been imposed on the color scaling in the right-hand panel for clarity.
		\textbf{Middle:} As above, but for intrinsic SED number (see \S\ref{subsec:som_rest}). The general substructure the underlying SEDs imprint on the SOM as a function of redshift is clearly visible (left), while trade-off between redshift accuracy and intrinsic SED uncertainty due to dropout sources at $z \gtrsim 5 $ can also be seen (right). An upper limit of $\textrm{std}(\textrm{SED})=50$ has been imposed on the color scaling in the right-hand panel for clarity.
		\textbf{Bottom:} As above, but for reduced-$\chi^2$. The observed variability as a function of observed-color in this case is a combination of sparse sampling of the model catalog, some limited intrinsic variability in galaxy colors, and the uneven color coverage our archetypes provide in certain regions of parameter space. An upper limit of $\textrm{std}(\textrm{reduced-}\chi^2)=10$ has been imposed on the color scaling in the right-hand panel for clarity.}
	\label{fig:3d_som}
\end{figure*}
\begin{figure*}
	\centering
	\captionsetup[subfigure]{labelformat=empty}
	\subfloat[][]{\includegraphics[scale=0.20]{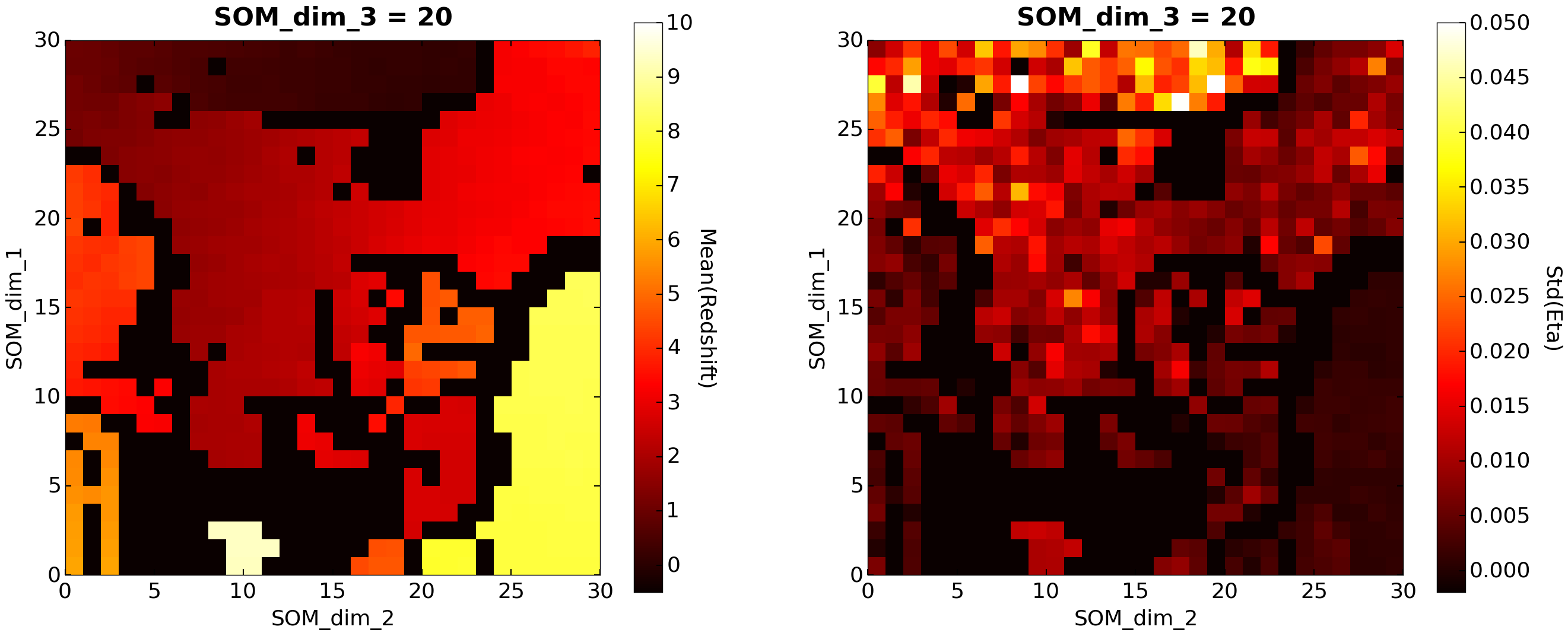}}
	\qquad
	\subfloat[][]{\includegraphics[scale=0.20]{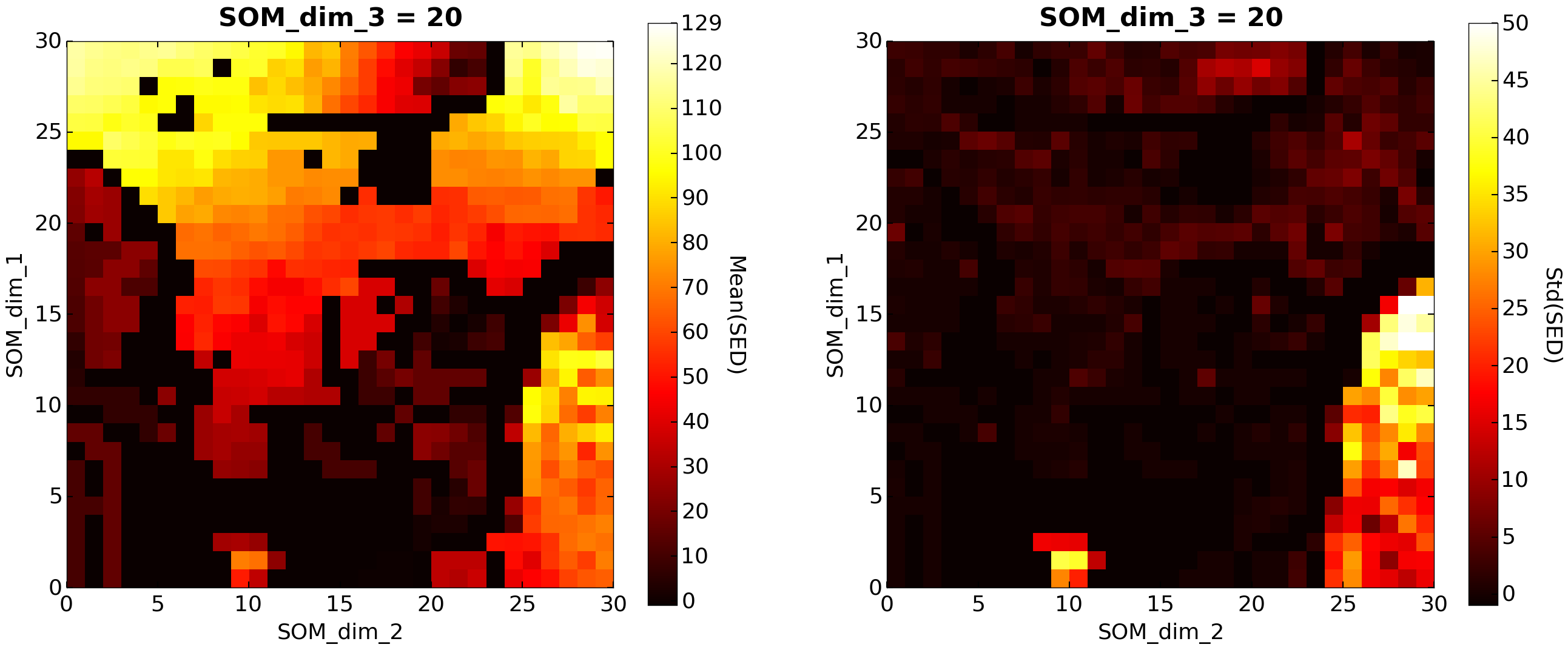}}
	\qquad
	\subfloat[][]{\includegraphics[scale=0.20]{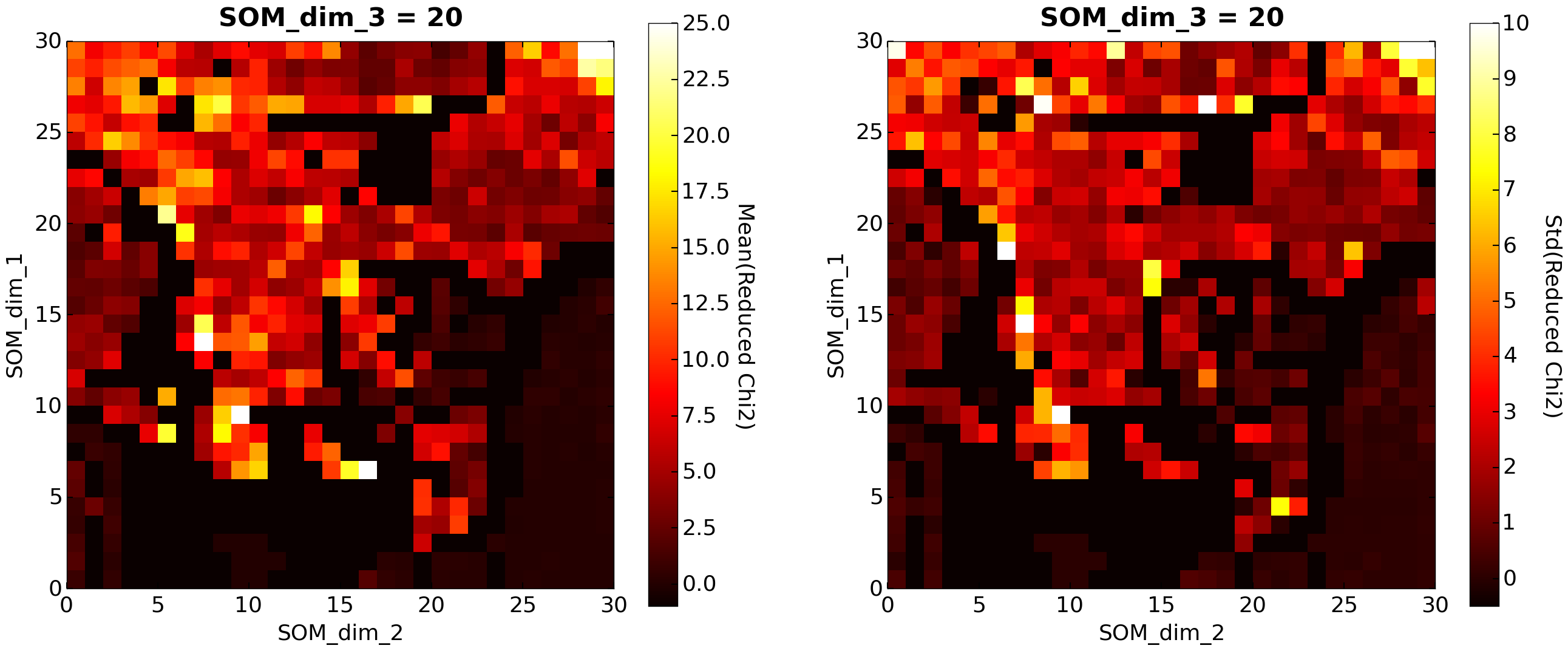}}
	\caption{
		As Figure~\ref{fig:3d_som}, but now illustrating a different 2-D slice. 
		\textbf{Top:} As the top panels on the previous page. Similar features are also present in this redshift slice (left), with greater uncertainty/degeneracy in the color-redshift relation at lower redshifts again visible in the right panel.
		\textbf{Middle:} As the middle panels on the previous page. Although the distribution of average SEDs across the map is markedly different than in the $x_3=10$ slice (left), the trade-off between redshift accuracy and intrinsic SED uncertainty due to dropout sources remains pronounced (right).
		\textbf{Bottom:} As the bottom panels on the previous page. While some of the variation is again due to sparse sampling, larger amounts of variation due to emission line contributions from bluer galaxies are now also present.}
	\label{fig:3d_som_2}
\end{figure*}

\begin{figure*}
	\centering
	\captionsetup[subfigure]{labelformat=empty}
	\subfloat[][]{\includegraphics[width=0.5\textwidth]{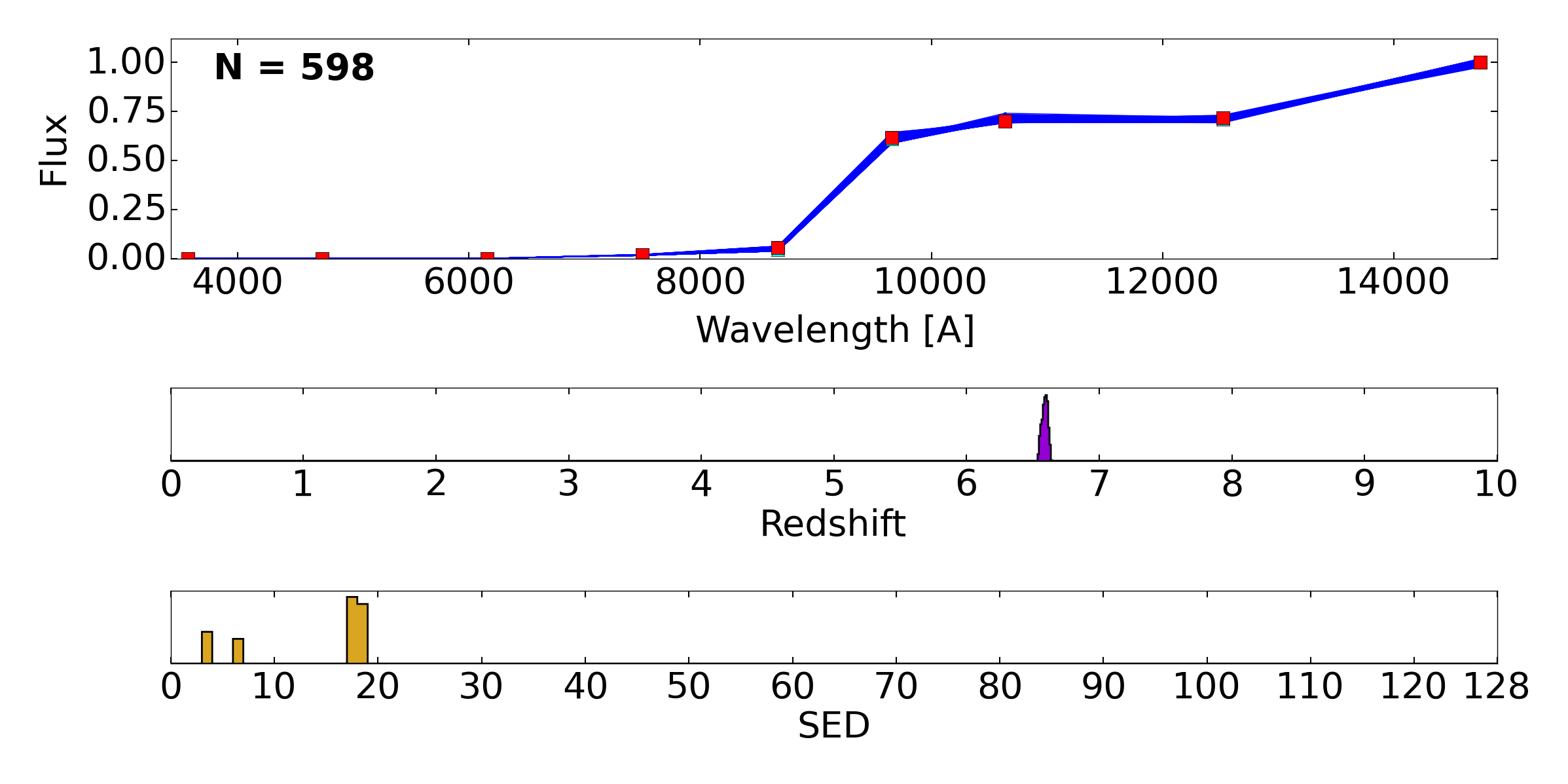}}
	\subfloat[][]{\includegraphics[width=0.5\textwidth]{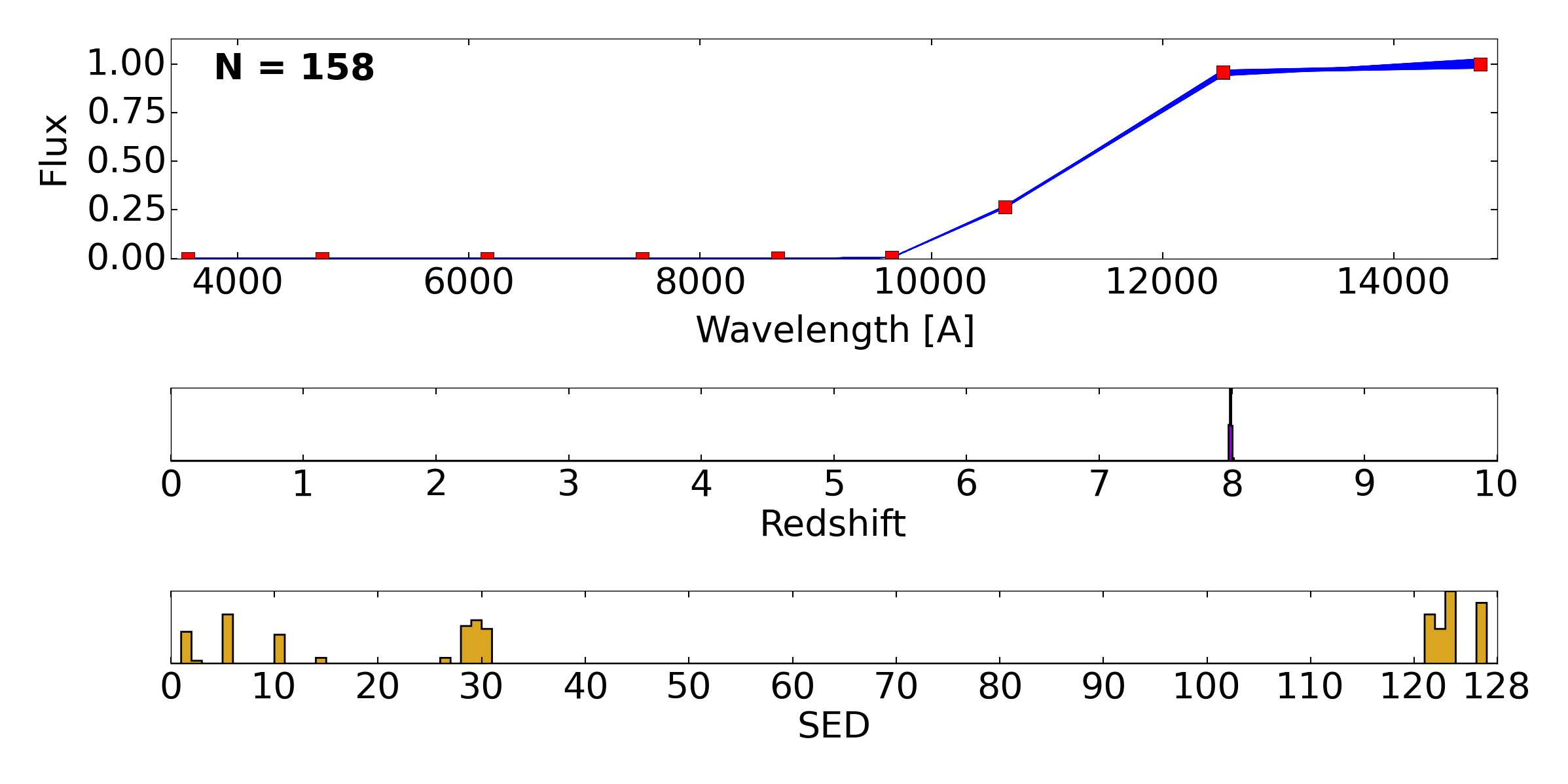}}
	\qquad
	\subfloat[][]{\includegraphics[width=0.5\textwidth]{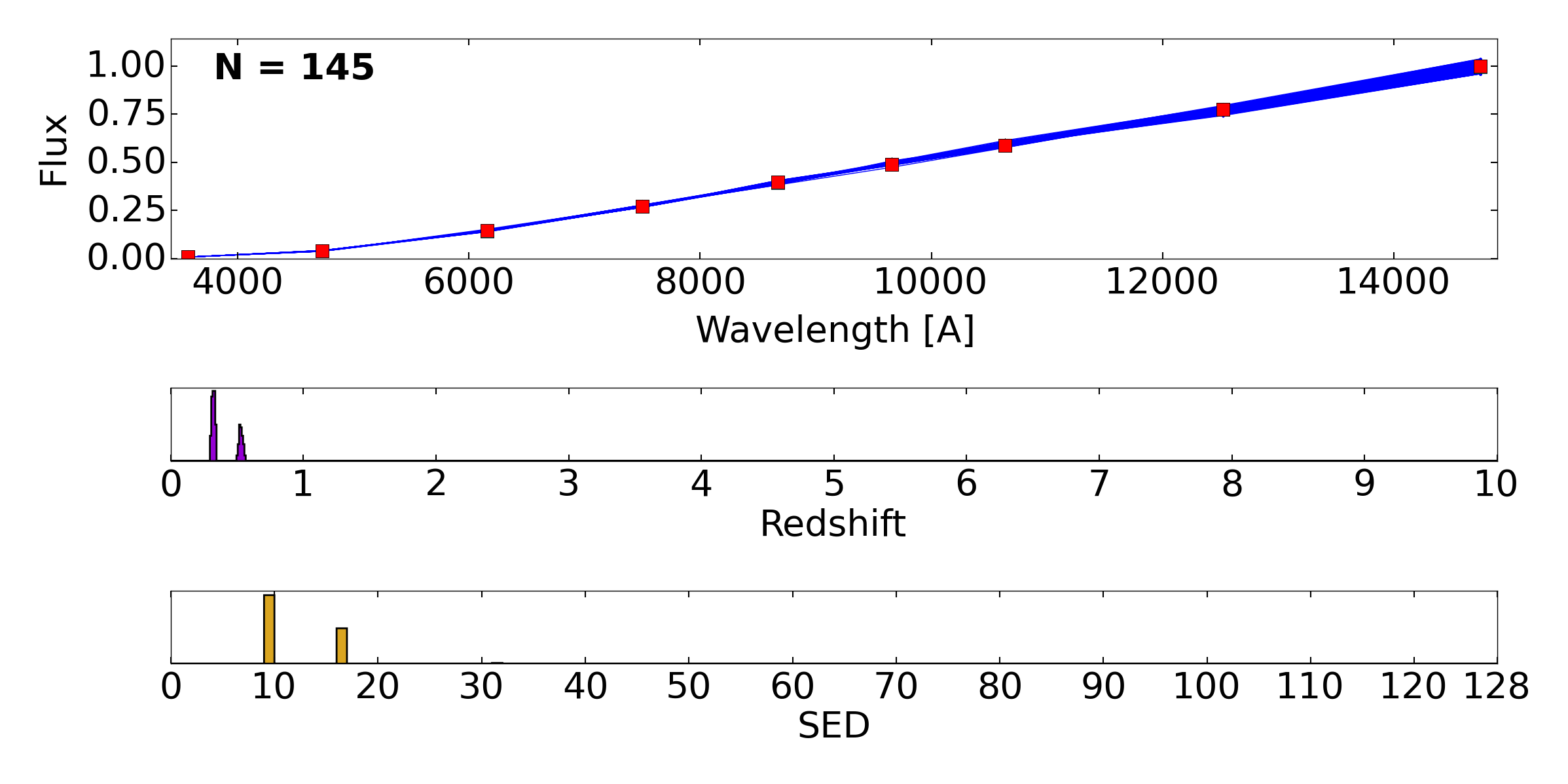}}
	\subfloat[][]{\includegraphics[width=0.5\textwidth]{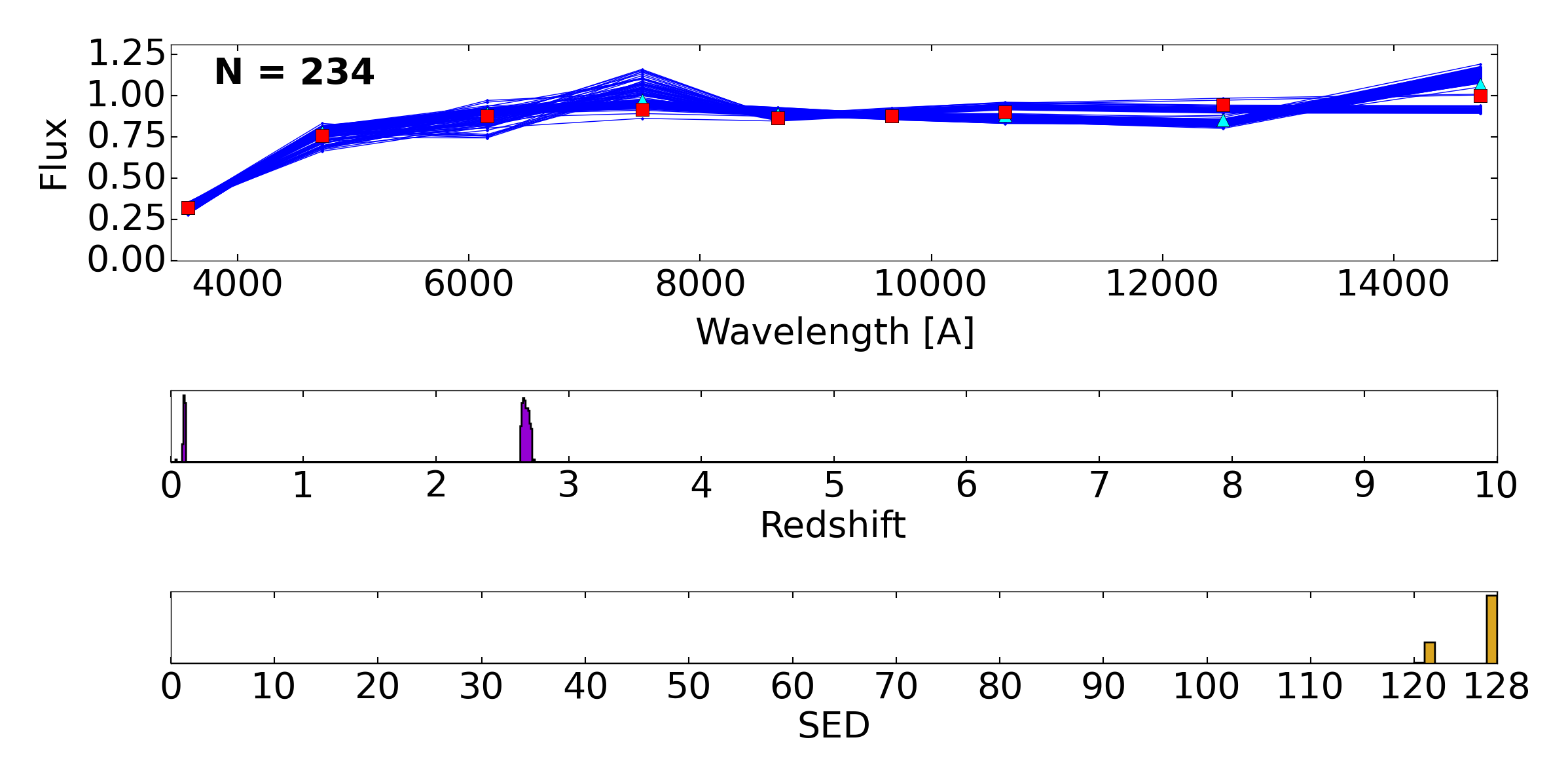}}
	\caption{
		Several individual cells taken from the 3-D observed-frame SOM shown in Figure~\ref{fig:3d_som}. The SOM cell models $\rom{\mathbf{F}}{som}^{\textrm{cell}}$'s are plotted in the upper portion of each panel as red squares along with their corresponding set of input model photometry (blue lines; means plotted as cyan triangles), with the associated redshift (purple) and intrinsic SED (yellow) distributions shown in the middle and bottom portions of each panel, respectively.
		\textbf{Top left:} Although the general redshift distribution in this cell is relatively tight, the intrinsic SED distribution turns out to be multimodal.
		\textbf{Top right:} The large number of dropout bands and the specific spectral shape redwards of $\sim$\,10,000\,{\AA} helps to ensure the redshift distribution in this cell is extremely narrow. However, this same information provides only weak constraints on the intrinsic shape of the SED.
		\textbf{Bottom left:} Two of our fuzzy archetypes produce similar photometry at two different redshifts, leading to an intrinsic bimodality in the color-redshift-archetype relation (i.e. the redshift-reddening degeneracy) in this particular region of color space. This bimodality would be difficult to capture using an exclusively forward-modeling approach, but is recovered naturally here.
		\textbf{Bottom right:} Although the amount of observed variation is significantly larger due to the emission line variability allowed by the blue star-forming archetypes captured in this cell, confusion between the 1216\,{\AA} and 4000\,{\AA} breaks in this specific region of color space leads to a widely-separated bimodal redshift distribution that is recovered by the SOM.}
	\label{fig:3d_som_cells}
\end{figure*}


\subsection{Application to Observed-Frame Photometry}
\label{subsec:som_obs}

We use the SOM to provide a similar topological re-organization to the entire observed-frame color space spanned by our fuzzy archetypes over a given redshift interval. As discussed in \S\ref{subsec:som_model}, this allows us to explore variations in observed-frame color directly when during the SED fitting process, while allowing the SOM to locate wormholes in the SED-redshift space formed by higher-dimensional combinations of parameters. Capturing these widely-separated degeneracies is crucial to properly constructing multi-modal $P(z)$'s.

In particular, a SOM constructed using observed-frame colors should be able to capture two key degeneracies:
\begin{enumerate}
\item \textit{Bimodal redshift PDFs}. Objects can contain widely-separated redshift degeneracies due to confusion between redshift and reddening or the 1216\,{\AA} and 4000\,{\AA} breaks, among others \citep{sobral+12,ilbert+13,steinhardt+14}. This leads to a multi-modal $P(z)$ that is difficult to fully capture without resorting to sophisticated sampling techniques \citep{feroz+13} or a brute-force grid search, both of which tend to be computationally expensive \citep{speagle+15}.
\item \textit{Complicated parameter degeneracies}. Different photometric bands convey varying amounts of information as a function of redshift. For instance, while an objects at high redshift can have a well-defined $P(z)$ based on a widespread lack of detections in a number of bands (the ``dropout'' technique), the same phenomenon often removes much of the information content about its underlying spectral shape, leading to weak constraints and/or large degeneracies on the intrinsic SED (and dust content) that are difficult to probe using an exclusively forward-modeling approach.
\end{enumerate}

As \S\ref{subsec:som_rest}, we create a model catalog for training using the fiducial set of fuzzy archetypes outlined in \S\ref{subsec:final_fuzzy}. For each of our 129 fuzzy archetypes, we generate 25 Monte Carlo realizations from $P(\phi)$ with uniform sampling from $\rom{k}{dust}$ for every point on an input redshift grid, which we take to range from $z=0$\,--\,$10$ with a redshift resolution of $\Delta z=0.01$. This gives us 3,228,225 sets of output model photometry to which we assign $1$\% errors as above. To include the effects of observational flux limits, we also impose an error floor of $0.01$ units of normalized flux, which we add to the errors in quadrature. These have mostly been chosen to guarantee acceptable performance since the corresponding model scale factors are unknown (i.e. we have no knowledge about the magnitude distribution as a function of $\boldsymbol{\theta}$ that our fuzzy archetypes templates should be drawn from). Ideally, however errors would be assigned according to the systematic uncertainties and observational limits of the survey itself, allowing the SOM to train using the \textit{specific knowledge content conveyed by particular survey(s) and/or dataset(s)}. We hope to investigate such modifications in future work.

Rather than using a 2-D SOM to probe observed-frame color space as in \citet{masters+15}, we instead construct a SOM using a three-dimensional grid of $30 \times 30 \times 30$ cells in order to allow the SOM additional freedom to organize the input data. While there is no intrinsic limit to the dimensionality of the SOM (ideally, the more dimensions are available the better the SOM can conform to the data distribution), we choose to restrict our map to only three dimensions to keep it relatively easy to visualize (which helps to assess performance). This assumption may be relaxed in future works as we develop more rigorous and/or abstract metrics to gage performance.

While our neighborhood function $\mathcal{H}(t)$ ensures that nearby cells are similar to each other (i.e. that the SOM is topologically smooth), it does not guarantee that the \textit{entire} SOM is a well-suited projection of the input dataset. In other words, there is no guarantee that individual regions on the SOM -- separated by some number of intervening cells -- will not be self-similar. This implies that the $\chi^2$ distribution of individual objects on the SOM might not be well localized, and the SOM will not give approximately unimodal solutions (within the manifold spanned by the SOM) for individual objects observed in all training bands.

This behavior is due in part to three separate issues: the limitations imposed by a small number of projected dimensions, the random initialization at the start of the training process, and the non-uniqueness of each projection. Since a number of projections from an $N$-dimensional space to an $M \ll N$-dimensional one often are equally ``good'' in terms of producing reasonable occupation rates and low $\chi^2$'s, for a random initialization of the SOM there is no particular preference as to which one will be chosen. While this behavior is acceptable in cases where the SOM is used purely as a clustering tool, it is a significant impediment when trying to use the SOM to visualize galaxy occupation of color space and makes the SOM-based sampling approaches described in \S\ref{sec:mcmc} significantly more difficult to implement.\footnote{Note that it is entirely possible that an object observed in only a small \textit{subset} of the training bands \textit{will} map to multiple regions on the SOM. This is due to information content lost from the missing bands and is a more difficult issue to bypass that we will not elaborate on here.}

As a result, rather than initializing our SOM cells randomly, we find we can improve the general performance of the SOM by \textit{seeding} the SOM cells using an appropriate series of color gradients. This gives it a rough starting topology form which it can later adapt, biasing the SOM to prefer certain projections over others based on our pre-existing knowledge of color space.

After experimenting with a variety of different seeding techniques and priors, we settle on using of a color grid in $u-r=[-0.5,5.3]$ ($\Delta (u-r) = 0.2$\,mag), $r-z=[-0.85,3.79]$ ($\Delta (r-z) = 0.16$\,mag), and $z-H=[-1.5,5.75]$ ($\Delta (z-H) = 0.25$\,mag). For each individual cell, we assign a starting normalization in the $u$ band of 1 and construct the cell model by linearly interpolating between bands using the defined color gradients at the corresponding cell position $\rom{\mathbf{x}}{som}^{\textrm{cell}}$ as a function of the effective wavelength in our corresponding filters.

After seeding our initial cell models, we assign our hyper-parameters $[a_0,a_1,\sigma_0]=[0.5,0.2,10]$ and allow the map to train for $\rom{N}{iter}=1,000,000$ iterations. These numbers were chosen after trial and error to maximize the utility of our seeding process by allowing the SOM to be less responsive (i.e. ``stiffer'') to new training data. This allows the SOM to adapt in a more perturbative fashion away from the initial seeded topology. After training, we map the 3,228,225 objects in our model catalog back onto the SOM using the corresponding set of best-matching cells. 

\begin{figure*}
	\centering
	\captionsetup[subfigure]{labelformat=empty}
	\subfloat[][]{\includegraphics[scale=0.16]{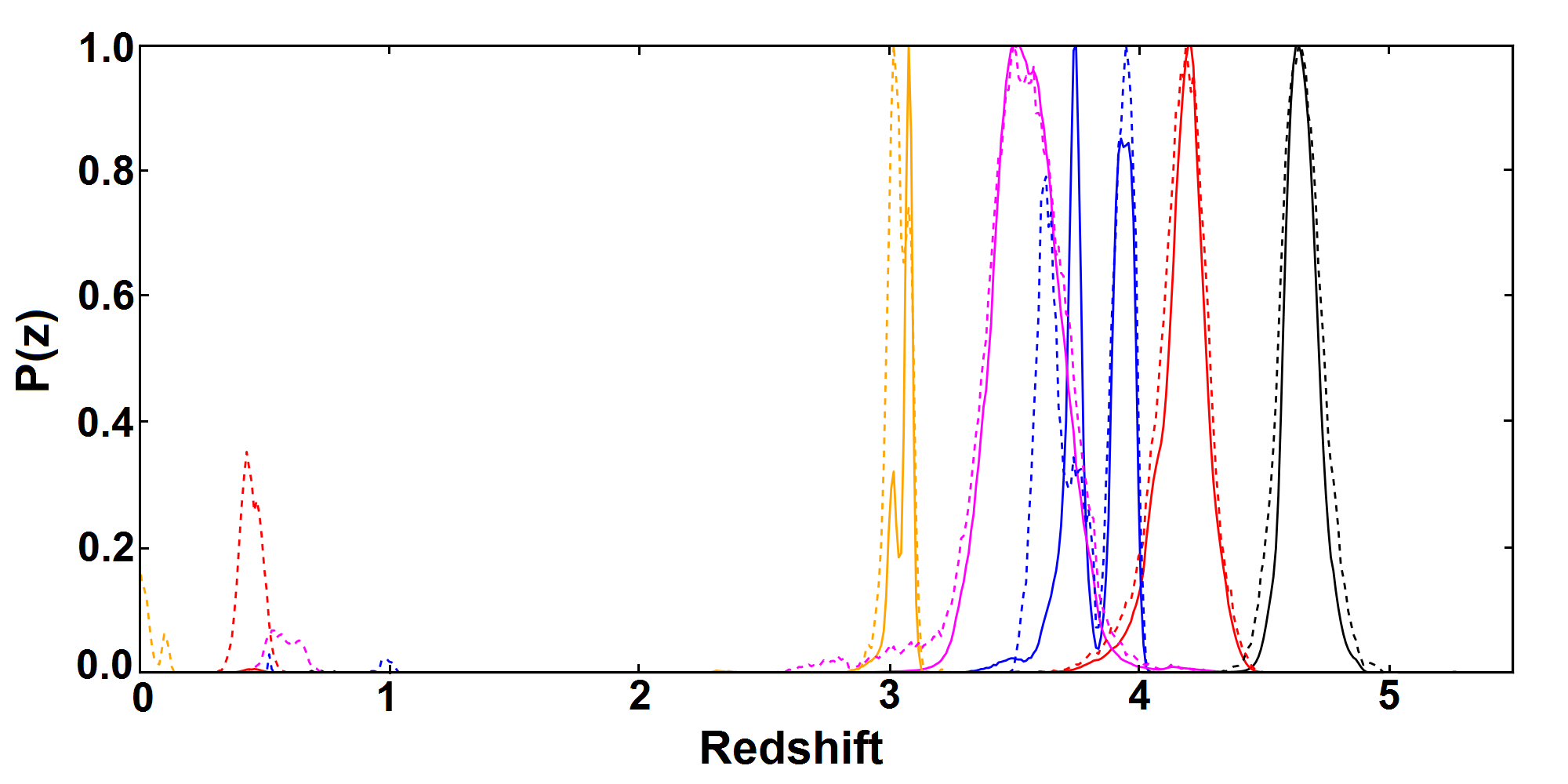}}
	\qquad
	\subfloat[][]{\includegraphics[scale=0.16]{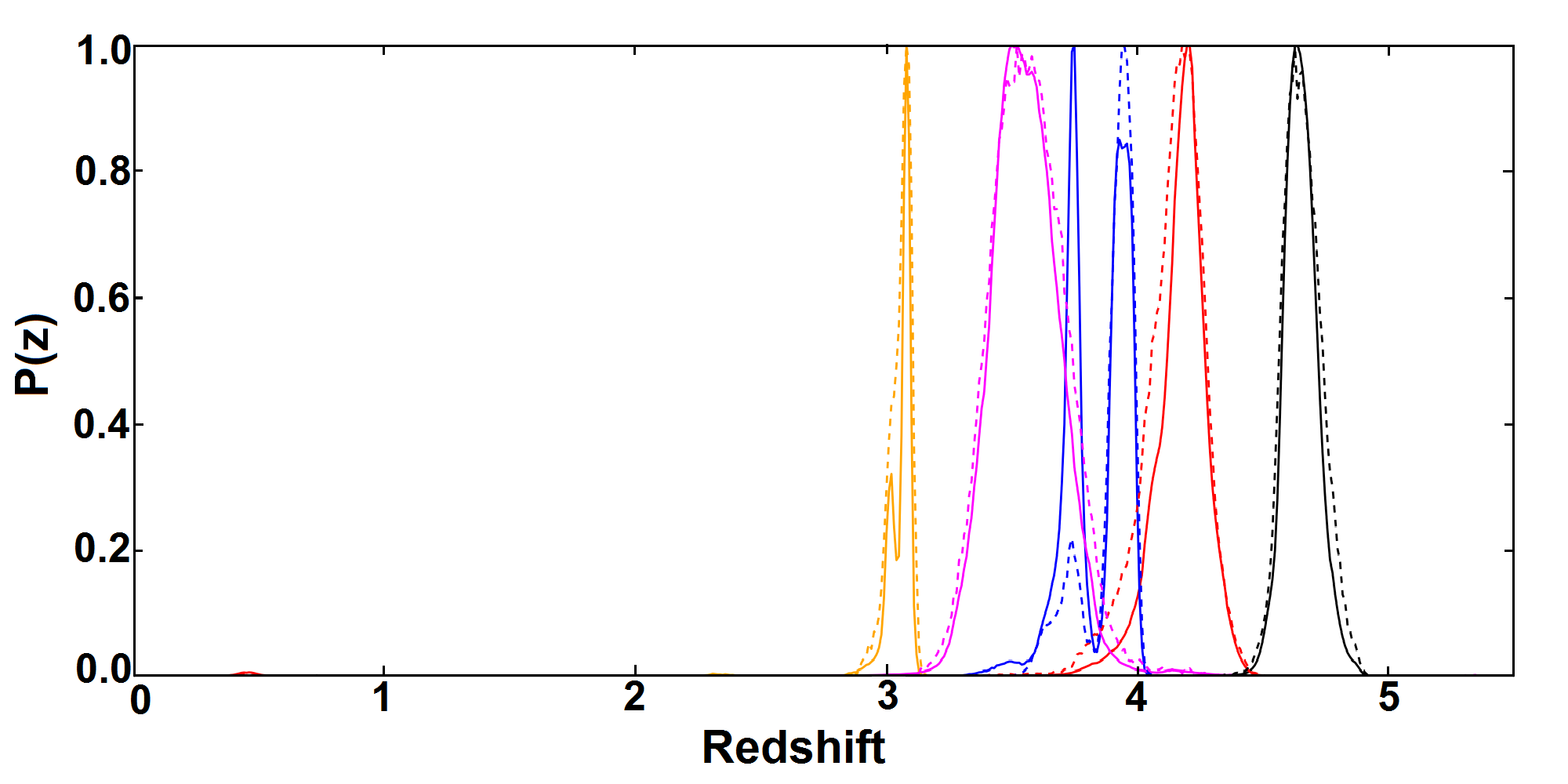}}
	\caption{
		The redshift probability distribution functions $P(z|\rom{\mathbf{F}}{obs})$ derived from a brute-force approach (solid lines) and the Markov Chain Monte Carlo (MCMC) 2+1-D rest-frame SOM+redshift approach outlined in \S\ref{sec:mcmc} (dashed lines) for five sample objects before (left) and after (right) the application of maximum likelihood-dependent weights.
		The unweighted $P(z|\rom{\mathbf{F}}{obs})$'s display a number of auxiliary peaks relative to the brute-force case due to chains becoming trapped in degenerate regions of color-redshift-archetype space (i.e. local minima) not fully removed by the 2-D rest-frame SOM. This behavior is effectively suppressed after weighting the contribution from each chain by their respective maximum likelihoods.}
	\label{fig:som_mcmc_pdf}
\end{figure*}

Two separate 2-D slices of our final 3-D observed-frame SOM, colored according to the mean and standard deviation of the $z(\rom{\mathbf{x}}{som}^{\textrm{cell}})$, $\textrm{SED}(\rom{\mathbf{x}}{som}^{\textrm{cell}})$, and reduced-$\chi^2(\rom{\mathbf{x}}{som}^{\textrm{cell}})$ distribution in a given cell, are shown in Figures~\ref{fig:3d_som} and~\ref{fig:3d_som_2}. While certain boundary regions demarcating large changes in color on the SOM remain unoccupied, the general occupation rate of the SOM (i.e. the fraction of cells which have $N(\rom{\mathbf{x}}{som}^{\textrm{cell}})>0$ objects) is relatively high at around $\sim 80\%$. 

In general, the observed-frame SOM is able to capture the observed color-redshift relation extremely well, with a well-defined evolution in the mean $z(\rom{\mathbf{x}}{som}^{\textrm{cell}})$ distribution across the map (top left panels) and a number of corresponding redshift wormholes (top right panels). In addition, it also correctly maps the limited information content and intrinsic degeneracies contained in the color-intrinsic SED relation across observed color space (middle panels) as well as the general variability in the color-redshift relation as a function of color (bottom panels), as showcased in Figure~\ref{fig:2d_som_2}. Several individual SOM cells and their corresponding distribution of input model photometry, redshifts, and SEDs are shown in Figure~\ref{fig:3d_som_cells}.


In summary, we find judicious use of the SOM involving multidimensional color seeding, error thresholds, and more ``perturbative'' training is effective at reorganizing the observed-frame color distribution of our fuzzy archetypes over a wide redshift range. In three dimensions, the SOM is able to capture the variable information content present in observed photometry, widely seperated degeneracies in redshift, and varying constraints on galaxy SEDs and other input parameters for a large collection of model photometry. This makes it well-suited to be used as part of a method that directly probes observed-frame color space to derive $P(z)$.

\section{Exploring Self-Organizing Maps Using Markov Chain Monte Carlo}
\label{sec:mcmc}

By itself, the SOM's ability to effectively cluster data is important -- good clustering algorithms enable the use of hierarchical sampling approaches that can jump from cluster to cluster, exploring both cluster-aggregated data properties as well as individual objects. What differentiates the SOM from other clustering algorithms, however, is that it also naturally defines the relative ``distance'' to each cluster in a straightforward manner based on the input cell grid. These combined properties make the SOM particularly amenable to sampling approaches that depend on general properties of smoothness and local gradients and/or that utilize discrete Markov chains to explore systems in a stochastic manner.

To take advantage of this, we turn to Markov Chain Monte Carlo (MCMC) sampling. Unlike brute-force approaches, which sample evenly and apply likelihood-dependent weights afterwards, MCMC-based algorithms \citep[see, e.g.,][]{johnson+13,foremanmackey+13,speagle+15} instead reconstruct the PDF by sampling \textit{proportional} to the likelihood in a stochastic manner and weighting all accepted trials evenly \citep[although see][]{bernton+15}.

A standard search heuristic employed by most MCMC codes for exploring an arbitrary set of parameters $\mathbf{\Theta}$ is the Metropolis-Hastings algorithm \citep{metropolis+53,hastings70}:
\begin{enumerate} 
	\item Draw a set of trial parameters $\mathbf{\Theta}_t$ at a given iteration $t$ from the corresponding \textit{proposal distribution} $\pi(\mathbf{\Theta}_{t-1}\rightarrow\mathbf{\Theta}_t)$.
	\item Accept the new trial and move to location $\mathbf{x}_t$ with a \textit{transition probability} $\alpha = \min \left( 1,\frac{P(\mathbf{\Theta}_t|\rom{\mathbf{F}}{obs})}{P(\mathbf{\Theta}_{t-1}|\rom{\mathbf{F}}{obs})}\frac{P(\mathbf{\Theta}_{t-1}\rightarrow\mathbf{\Theta})}{P(\mathbf{\Theta}_{t}\rightarrow\mathbf{\Theta}_{t-1})} \right)$. Otherwise, remain at $\mathbf{\Theta}_{t-1}$.
	\item Repeat from step (i) until a stopping criterion is reached.
\end{enumerate}
This procedure is used to guide several individual ``chains'' of related draws as they converge to (``burn in'') and eventually begin sampling from the region of interest. 

We implement a version of MCMC sampling that transitions to new regions based on comparisons between individual sets of model photometry but with chains that live ``on top of'' the SOM. Although we test a variety of proposal distributions, we ultimately settle on the following four-part proposal:
\begin{enumerate}
	\item $\pi(\rom{\mathbf{F}}{model}^{\textrm{cell}(t-1),n}\rightarrow\rom{\mathbf{x}}{som}^{t-1})=\rom{\mathbf{x}}{som}^{t-1}$: Transition from the previous model located in a given corresponding cell at iteration $t-1$ with index $1 \leq n \leq N(\rom{\mathbf{x}}{som}^{\textrm{cell}(t-1)})$ to the corresponding cell position $\rom{\mathbf{x}}{som}^{\textrm{cell}(t-1)}$.
	\item $\pi(\mathbf{x}_{\textrm{som}}^{\textrm{cell}(t-1)}\rightarrow \mathbf{x}_{\textrm{trial}}^{\textrm{cell}(t)}) = \exp\left[-\frac{\left(\mathbf{x}_{\textrm{trial}}^{\textrm{cell}(t)} - \mathbf{x}_{\textrm{som}}^{\textrm{cell}(t-1)}\right)^{\mathbf{2}}}{2\boldsymbol{\sigma}^\mathbf{2}}\right]$: Propose the position of the next cell at iteration $t$ by drawing from a series of independent Gaussian kernels, where $\mathbf{2}=2\mathbf{I}$ indicates element-wise exponentiation.
	\item $\pi(\mathbf{x}_{\textrm{trial}}^{\textrm{cell}(t)}\rightarrow \mathbf{x}_{\textrm{som}}^{\textrm{cell}(t)})=\textrm{round}\left(\mathbf{x}_{\textrm{trial}}^{\textrm{cell}(t)}\right)$: Transition from the proposed position to the corresponding SOM cell by rounding each component to the nearest integer.\footnote{Cases were the proposed position is outside the grid of cells are wrapped around to the opposite side.}
	\item $\pi(\rom{\mathbf{x}}{som}^{\textrm{cell}(t)}\rightarrow\rom{\mathbf{F}}{model}^{\textrm{cell}(t),m})=\textrm{randint}[1,N(\rom{\mathbf{x}}{som}^{\textrm{cell}(t)})]$: Transition to a random model with index $1 \leq m \leq N(\rom{\mathbf{x}}{som}^{t})$ contained within the given cell by drawing a random integer from $1$ to $N(\rom{\mathbf{x}}{som}^{t})$ (inclusive).
\end{enumerate}
While the formalism is slightly messy, this proposal is relatively simple to implement in practice since previous models and their corresponding cell locations can be stored in memory (thereby bypassing step one) and generating random numbers is straightforward.

Our proposal distribution, while symmetric in SOM space (steps (ii) and (iii)), is asymmetric in the target model space since the likelihood of selecting a particular model is inversely proportional to the number of models sorted into a given cell (i.e. the size of the corresponding cluster). The modified transition probability is
\begin{equation}
\rom{\alpha}{mod} = \min \left( 1,\frac{N(\rom{\mathbf{x}}{som}^{\textrm{cell}(t)})P(\rom{\mathbf{F}}{model}^{\textrm{cell}(t),m}|\rom{\mathbf{F}}{obs})}{N(\rom{\mathbf{x}}{som}^{\textrm{cell}(t-1)})P(\rom{\mathbf{F}}{model}^{\textrm{cell}(t-1),n}|\rom{\mathbf{F}}{obs})}\right),
\end{equation}
which satisfies detailed balance.

As noted above, portions of the SOM have cells that contain no sets of model photometry. To accommodate these regions of ``missing data'', we simply modify our original proposal distribution by repeating the cell proposal described in steps (ii) and (iii) above until $N(\rom{\mathbf{x}}{som}^{\textrm{cell}(t)})>0$. This allows chains to explore areas past these positions in a random-walk manner. As this proposal is symmetric, it does not necessitate any modifications to our acceptance probability and does not violate detailed balance.

We implement both 3-D (observed-frame SOM) and 2+1-D (rest-frame SOM+redshift evolution) versions of this sampling scheme. In the 3-D case, we run our MCMC procedure over the 3-D observed-frame SOM (Figure~\ref{fig:3d_som}) with a fiducial value of $\sigma_1=\sigma_2=\sigma_3=2$ cells for the SOM proposal distribution in each dimension. For the 2+1-D case, we only implement this sampling scheme on the 2-D rest-frame SOM with the same fiducial values, while redshifts are instead drawn from a separate Gaussian proposal distribution with $\sigma_z=2.5\%$ of the corresponding redshift grid (in this case $\sigma_z=0.25$).

\begin{figure*}
	\centering
	\captionsetup[subfigure]{labelformat=empty}
	\subfloat[][]{\includegraphics[width=\linewidth]{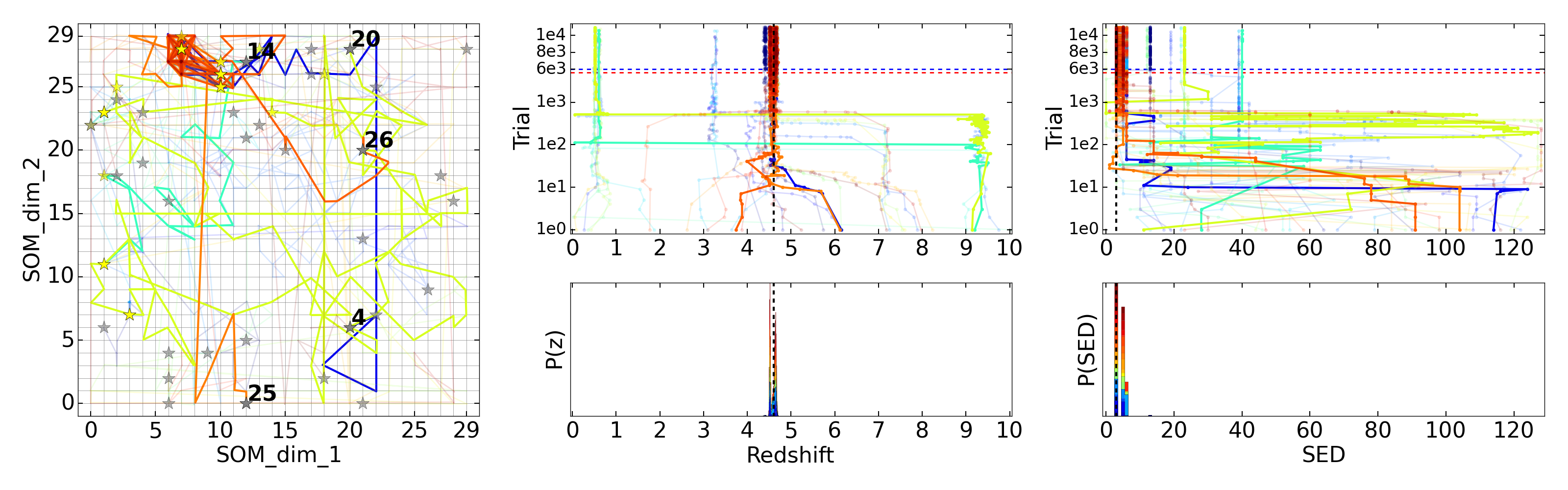}}
	\qquad
	\subfloat[][]{\includegraphics[width=\linewidth]{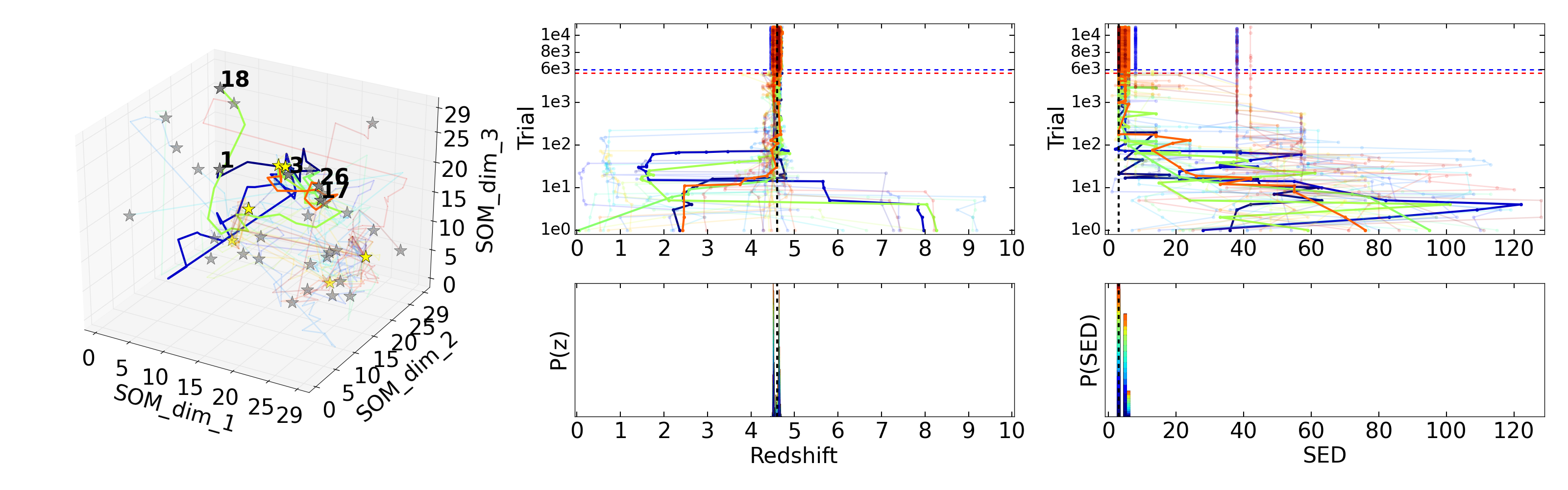}}
	\caption{Individual runs for a mock object (a red elliptical template located at $z \sim 4.5$) sampled using 32 chains over our 2+1-D rest-frame SOM+redshift (\textbf{top}) and 3-D observed-frame SOM (\textbf{bottom}). In each plot, the general evolution over the course of the run of the positions of the individual chains on each SOM (left), in redshift (middle), and in intrinsic SED (right) are shown, with a random set of five individual chains highlighted. Only accepted trials are plotted for clarity. The initial starting positions of each of the chains are indicated as gray stars while their maximum likelihood positions are marked as yellow stars. The evolution of redshift and intrinsic SED are plotted logarithmically during the burn-in (below the dashed red line) and learning phases to illustrate rapid evolution in chain positions, and in linear space afterwards (above the dashed blue line) to draw attention to their final stationary behavior. The weighted marginalized PDFs (see Figure~\ref{fig:som_mcmc_pdf}) of both parameters are also shown, with dashed black lines over-plotted to indicate their true values. While in general the majority of chains converge to and sample from the correct regions of parameter space, the explicit dependence on redshift in when using the 2-D rest-frame SOM creates a broad redshift-dependent gradient that can lead chains to degenerate regions of parameter space with only marginal likelihoods, impeding overall convergence. These issues are mostly resolved in the full 3-D observed-frame case where all 32 chains converge to the correct redshift distribution.}
	\label{fig:som_mcmc_verbose}
\end{figure*}

To improve performance and reduce the dependence on our initial choice of parameters, we allow our simple MCMC sampler a limited ``learning step'' during every MCMC run after the burn-in phase is complete. During this step, we draw samples from the fiducial neighborhood function for a set number of iterations, compute their corresponding standard deviations, and use them to update the relevant scale of the fiducial neighborhood function in each dimension. This allows us to adjust our neighborhood function to the approximate the shape of the PDF and tune the general acceptance fraction in order to improve subsequent performance.

While the SOM in general is topologically smooth, we find that its re-oganization of the full parameter space does not necessarily remove all local minima for individual objects. This is especially true for our 2+1-D implementation, which by construction will still encounter widely-separated redshift degeneracies. In order to deal with the fact that different chains will likely converge to and sample from any number of individual minima, we assign each chain a given weight based on the maximum likelihood found during the course of the its run. This is then used to reconstruct the total probability probed by all the chains via
\begingroup\makeatletter\def\f@size{8.5}\check@mathfonts
\begin{equation}
P(z|\rom{\mathbf{F}}{obs})=\sum_{\textrm{chain}} \textrm{max} \left(\left\lbrace P(\rom{\mathbf{F}}{model}|\rom{\mathbf{F}}{obs}) \right\rbrace_{\textrm{chain}}\right) P(z|\textrm{chain},\rom{\mathbf{F}}{obs}).
\end{equation}
\endgroup

The general effectiveness of this scheme to improve naive $P(z|\rom{\mathbf{F}}{obs})$ estimates derived using the 2+1-D SOM+MCMC algorithm with 32 chains as compared to a brute-force counterpart in shown in Figure~\ref{fig:som_mcmc_pdf}. While this weighting scheme effectively suppresses the contribution from chains that converge to minima which are significantly worse than the global minimum, it is less effective at dealing with minima with have comparable likelihoods, where initial conditions and sampling variance (i.e. how many chains converge to a given location) dominate the final PDF. Although this scheme is not perfect, ours tests indicate it is sufficient for our purposes.

To illustrate the general behavior of our hybrid SOM+MCMC algorithms, visualizations of an specific run for an inidividual object is plotted in Figure~\ref{fig:som_mcmc_verbose}. The object is generated from our collection of fuzzy templates at a given redshift, normalized to $24$\,mag in the LSST $Y$-band, and given error properties according to the expected observing depths of each survey as described in \S\ref{subsec:som_model}. In each plot, the general evolution over the course of the run of the positions of the individual chains on the SOM (2-D vs. 3-D), their redshift (an explicit parameter in the 2-D case; an implicit one in the 3-D case), and their intrinsic SED (implicit in both cases) are shown, with a random set of individual chains highlighted. 32 chains were run in parallel with 5000, 1000, and 5000 iterations allocated for burn-in, learning, and sampling the stationary distribution, respectively.

While the 2-D rest-frame SOM effective re-organizes the rest-frame template space (Figure~\ref{fig:2d_som_2}), the overall non-linear effect of redshift evolution imposes a general gradient on the full 2+1-D space that guides individual chains to specific redshift modes depending on their initial redshift. This constraint also translates into a restriction on the amount of SED variation that can be easily explored: while underlying SEDs in separate regions of the map might look similar for a particular set of input photometry at a specific redshift, the random-walk nature of our MCMC chains cannot easily traverse these large distances and as a result cannot easily explore the available parameter space.

Both of these problems are alleviated in our 3-D observed-frame SOM+MCMC implementation. While the 3-D SOM doesn't necessarily eliminate \textit{all} spurious minima, we find it removes the strong dependence of the behavior of individual chains on their initial redshift positions and more generally improves convergence to the region of interest. It also grants additional flexibility when exploring larger variations in the underlying SED.

\section{Discussion}
\label{sec:disc}

At this point, we have established the general ingredients behind our data-driven, template-fitting approach, and demonstrated its ability to effectively recover redshift PDFs using MCMC. We will now discuss some of the implications of our general methodology, including its relationship to machine-learning approaches, observational requirements, computational improvements, and alternate statistical interpretations.

\subsection{Relationship to ``Empirical'' Machine Learning Techniques}
\label{subsec:ml_relationship}

Throughout this paper, we have emphasized that our archetype-driven approach is a relatively ``empirical'' one, especially when compared to traditional template-fitting approaches. However, we wish to spend some time distinguishing between the \textit{semi}-empirical nature of an archetype-driven approach and the \textit{fully}-empirical nature of machine-learning approaches.

The first main aspect of note about our approach is that the inherent ``fuzziness'' of the chosen archetypes is a tunable feature that fundamentally serves a way to (linearly) extend templates into neighboring regions of color space. As the corresponding coverage of color space becomes denser and/or more representative, less fuzziness will be required. As a result, as the amount of ``training data'' we are able to take advantage of increases (giving improved model catalogs and SOM projections), the corresponding model dependencies will likewise decrease, rendering our approach increasingly empirically-driven.

However, although we are empirically sampling the galaxy population, our approach is still fundamentally a template fitting-driven one because it relies on the assumption that information obtained at one redshift can be translated into other redshifts \textit{in a particular way} (as outlined in \S\ref{subsec:gen_phot}). Such an assumption is intuitive and reasonable, and allows us to leverage our physical understanding of how galaxies evolve over time.

By contrast, supervised machine-learning approaches make no such assumption. Instead, all relationships are derived entirely from the training data based on a particular chosen set of functional forms and methods to partition the corresponding color space. As a result, while machine-learning algorithms can give accurate predictions, they are not able to take advantage of this underlying continuity of information between one redshift and the next. While this does not prevent them from being remarkably effective \citep[see, e.g.,][]{hildebrandt+10,sanchez+14}, we believe the assumption of information continuity across redshifts is a key strength of the archetype approach that gives the method additional predictive power and a more intuitive level of understanding.

\subsection{Observational Requirements}
\label{subsec:req_obs}

As mentioned in \S\ref{subsec:phot_trad} and \S\ref{subsec:phot_archetype}, one of the largest drawbacks of template-fitting approaches is the observationally-demanding nature of collecting good templates. This is dependent on two main issues:
\begin{enumerate}
	\item Photometric surveys generally have (internally) well-calibrated photometry that is accurate to $\lesssim$\,$5$\% ($\lesssim$\,$0.05$\,mag uncertainty). Templates must be flux-calibrated to similar levels or better in order to avoid having observational systematics dominate $P(z)$, which requires very high S/N spectrophotometric observations and sophisticated flux calibration techniques (Ben Johnson et al., in preparation).
	\item In addition, surveys usually have broadband coverage that exceeds the range of any individual spectrograph. Templates thus must be stitched together from multiple observations and/or extended into regions with no coverage using stellar population synthesis models.
\end{enumerate}

The \citet{brown+14} templates used in this work goes to great lengths to deal with both of these issues using spectra taken at multiple wavelengths (FUV, optical, IR) with accurate flux calibration and careful extrapolation to other wavelengths using stellar population synthesis code \texttt{MAGPHYS} \citep{dacunha+08}. However, their coverage of parameter space is limited, with notable gaps in regions of high star-forming activity (see Figure~\ref{fig:brown_templates}). This is not unique to the \citet{brown+14} dataset -- almost all existing template sets \citep{coleman+80,kinney+96,polletta+07} suffer from the same gap, which is often filled in exclusively using stellar population synthesis models \citep[see, e.g.,][]{ilbert+09}. This gap is due to the limited redshift range probed by current templates, which only probes the nearby universe and thus necessarily misses the markedly different galaxy types present at higher redshifts.

For our archetype-driven method and others like it to be effective, this gap must be filled. This will require a deep observational campaign to get high S/N spectrophotometry of higher-redshift sources along with their emission lines strengths. We advocate for such a campaign focusing on galaxies around the ``sweet spots'' at $z \sim 0.9$ and $z \sim 2$, where the majority of spectral features are still present in the optical/NIR window. At $z \sim 0.9$ redshift, several major emission lines (H$\alpha$+N{\scriptsize{[II]}}, H$\beta$, O{\scriptsize{[III]}}, etc.) are located  in the $i$, $z$, and $J$ bands, while at $z \sim 2$ these mostly fall between $J$, $H$, and $K$. Optical and NIR spectroscopy accompanied by deep photometric observations of these targets to achieve a S/N ratio of $\gtrsim$\,50 should be doable, if difficult. Such follow-up spectroscopy could be done at ground-based facilities using instruments as MMIRS (Magellan/MMT) or NIRSPEC/MOSFIRE (Keck), or using space-based facilities using JWST's NIRSpec.

In addition to a deep spectrophotometric campaign to fill in this gap in parameter space, we might also supplement integrated spectroscopy of entire galaxies with templates derived from individual regions observed with integral field spectroscopy. While high-redshift galaxies as a whole might not resemble low-redshift galaxies, they might in fact resemble particular \textit{regions} of those galaxies such as, e.g., dusty star-forming clumps. Rigorous spectrophotometric fitting of data from integral field spectroscopic surveys such as the Calar Alto Legacy Integral Field Area survey \citep[CALIFA;][]{sanchez+12} and the SDSS \citep[Sloan Digital Sky Survey;][]{york+00}-IV Mapping Nearby Galaxies at Apache Point Observatory survey \citep[MaNGA;][]{bundy+15} using additional FUV/NIR photometric information could provide a whole new range of possible archetypes. We are currently in the preliminary stages of exploring enhanced spectrophotometric fitting techniques that should enable both of these approaches.

\subsection{Optimizing Exploration of the SOM}
\label{subsec:comp_improvements}

As showcased in \S\ref{sec:mcmc}, our MCMC-driven approach to exploring model photometry using the SOM is able to accurately reconstruct $P(z)$. However, it is far from optimized. As briefly discussed in \citet{speagle+15}, modern computing architectures are efficiently pipelined, enabling the use of vectorized operations and effective memory allocation for processes without any conditional statements. As a result, many repeat operations that reference memory in a predictable way can often be done not only in parallel using GPUs but with extremely efficient usage of the cache. Such performance is crucial for codes with essentially instantaneous likelihood allocations (i.e. with pre-generated photometry), where ``cache misses'' can lead to an order of magnitude increase in fetching time. This is enables optimized brute-force codes to run significantly faster than adaptive sampling techniques such as MCMC, even if the latter involves an order of magnitude or more fewer operations.

Due to the significant amount of random numbers generated in our simplistic approach outlined in \S\ref{sec:mcmc} (picking the cell, selecting a model, and deciding whether to transition to the model), our fiducial method is particularly susceptible to cache misses. To illustrate some of the qualities a more efficient sampling algorithm might possess, we briefly outline a three-tiered approach that makes better use of the cache and the localized positions of each chain in SOM space. 

Rather than computing likelihoods for individual $\rom{\mathbf{F}}{model}$'s, we could instead transition to computing cell-based quantities such that
\begin{equation}
P(\rom{\mathbf{x}}{som}^{\textrm{cell}}|\rom{\mathbf{F}}{obs})=\sum_i P(\rom{\mathbf{F}}{model}^{\textrm{cell},i}|\rom{\mathbf{F}}{obs}),
\end{equation}
where the sum, a corresponding flag, and all the individual model likelihoods would then stored in memory. As this calculation is a straightforward, vectorizable operation, it can likely be done extremely quickly. Chains would then transition according to these cell probabilities, which would be stored in memory along with the resulting accepted cell locations. Instead of storing these individually, we could simply count the number of times a chain visits each cell, such that after the run has completed, the final redshift distribution could then be constructed through a weighted sum of the individual $P(z|\rom{\mathbf{x}}{som}^{\textrm{cell}},\rom{\mathbf{F}}{obs})$'s. To ensure efficient burn-in, we could construct an initial ``first pass'' by computing a brute-force comparison across the individual \textit{SOM cell models} (with possible contributions from the variance in a given cell) -- essentially ``triaging'' the SOM likelihood surface -- after which each chain could be initialized in particular locations on the SOM that are likely to contain higher likelihoods. Finally, we could sample randomly from the full collection of computed model likelihoods (based on the frequency with which they were sampled) to generate a series of draws from the respective chain(s) if/when necessary.

While this procedure might not be the most effective way to achieve speedup, it does illustrate the computational flexibility calculations using the SOM provide. Given the efficiency of pipelines computers now possess to manage memory usage, optimization procedures like this are crucial to establish these types of approaches as computationally competitive with machine-learning implementations. This is especially true with our dense but discrete realization of the template set, which allows us to avoid recomputing likelihoods for templates that vary only insignificantly from past computations by caching and reusing past results.


\subsection{Alternate Statistical Interpretations: Exploring Graphs versus Hierarchical Sampling}
\label{subsec:discrete_markov_chains}

Clustering methods such as the SOM represent a unique form of dimensionality reduction that transforms a number of discrete realizations of a continuous (or at least effectively continuous) distribution of data into a finite number of clusters. Exploring such a discrete space is a challenging problem that has received much attention in statistics, often in the form of exploring ``graphs'' composed of ``nodes'' (clusters) and ``edges'' (connections between clusters). One of the most effective forms of exploring these graphs to derive structure and infer parameters is through the use of discrete transition MCMC methods that can jump from node to node, allowing chains to explore the graph through a (biased) random walk.

Much of this power comes from the flexibility of the proposal distribution (see \S\ref{sec:mcmc}). Unlike other minimization techniques that require continuous spaces and defined gradients, Markov chains can directly take advantage of discrete proposal distributions when attempting to transition from one region to the next as long as the corresponding transition probability is modified accordingly to satisfy detailed balance. While in our simplistic implementation above we have used the discrete realization of a continuous Gaussian distribution, we also experimented with several discrete proposal distributions such as uniform cell ``blocks'' that gave comparable performance.

One of the benefits (and limitations) of the SOM is that it defines a distance between every single cell throughout the training process (i.e. it establishes edges between every node along with an appropriate distance weight), allowing us to work in a quasi-continuous space that is discretely sampled by the SOM cells. However, there is no reason why this approach can't be applied to other more flexible graph-based clustering approaches such as a Growing Neural Gas (\citealt{fritzke95}; see \citealt{hocking+15} for a recent application in astronomy).

While such implementations are useful, clustering methods more generally do not require such a smooth surface or connected graph to be effective. Instead, they can simply be viewed as methods to \textit{partition} data according to some quality criteria, regardless of its underlying distribution. These can be used to generate fully independent clusters such as with $k$-means clustering \citep{forgy65,lloyd82} or hierarchical structures through the use of $k$-$d$ trees \citep{bentley75}. Determining the PDF of any individual observed data point then is just a question of its probabilistic association with these individual clusters, either through the use of summary statistics or other more holistic criteria.

In essence, this transforms the idea of sampling individual model parameters directly to instead sampling hierarchical structures that encapsulate varying amounts of information about the properties of different \textit{groups} of model parameters. We have crudely used such an approach in the fiducial MCMC sampling method outlined in \S\ref{sec:mcmc} by sampling randomly from the objects contained within each cell on the SOM, a compromise that allows our individual chains to explore the distribution data within each cluster in a stochastic fashion while also exploring the general distribution of clustered data in a stochastic way. However, there is no intrinsic need to use MCMC-based approaches to explore such hierarchical structures, leaving open the possibility that future photo-z approaches building on our current methodology might be able to exploit the $\mathcal{O}(\log N)$ speedup often afforded by hierarchical clustering approaches in computationally efficient ways.

\section{Conclusion}
\label{sec:conc}

Fast and accurate photometric redshifts (photo-z's) are a key requirement for future large-scale surveys hoping to derive distances to the billions of galaxies in their datasets. While much progress has been made in recent years using machine learning to empirically determine the mapping between color and redshift \citep{carrascokindbrunner13,carrascokindbrunner14,bonnett15,hoyle15,almosallam+15,elliott+15}, template-fitting approaches still suffer from numerous systematic uncertainties and computational inefficiencies involved with generating and fitting large ''grids'' of pre-generated model photometry. Part of the issue stems from the difficulty in properly modeling the impact of dust and emission line variation on an often very limited set ($\lesssim$\,$30$) of templates in order to simulate regions of color space where we know galaxies exist, coupled with the brute-force approaches most template-fitting codes in use today take to exploring these model photometric grids \citep{speagle+15}.

We have attempted to tackle both of these issues in order to address some of these deficiencies that currently prevent template-fitting methods from scaling up to the ``big data'' regime, with the overriding goal of creating a general framework that can not only accommodate a large set of input templates with some (limited) model-dependent variation but also be explored in approximately constant time within reasonable memory constraints. Our main results are as follows:
\begin{enumerate}
	\item Using a set of baseline archetypes from \citet{brown+14} that sample the complicated color-space manifold in color space where galaxies live and evolve, we establish a framework for creating expanded ``fuzzy archetypes'' that allows for limited amounts of dust attenuation and emission line variation to be probed independently with minimal modeling assumptions.
	\item Based on a series of reasonable approximations, we show how these templates can be decomposed into a set of linear photometric components that can be efficiently stored in memory and marginalized over/optimized out during the fitting process.
	\item We illustrate the ability for Self-Organizing Maps (SOMs) to re-organize both rest-frame and observed-frame ``model catalogs'' generated from Monte Carlo sampling of a set of fuzzy archetypes. This includes the ability to capture widely-separated degeneracies among parameters (i.e. ``wormholes'' through color space), efficiently cluster the corresponding data, and create topologically smooth discrete representations of reduced-dimensional manifolds that can be explored through hierarchical sampling techniques.
	\item We demonstrate the ability for discrete transition Markov Chain Monte Carlo (MCMC) techniques to explore parameter space by sampling over cells on the SOM. Individual models contained within the cells that are explored by each chain can then be used to recover redshift probability distribution functions (PDFs) of individual objects.
	\item Finally, we discuss some of extensions of our general approach, including observational requirements and opportunities, computational optimization, and possible applications to broader clustering-based methods.
\end{enumerate}

Much of the work in this paper has gone towards establishing the general framework and methodology of our approach. As a result, it should be only be viewed as the first tentative steps -- in the same vein as \citep{speagle+15} -- to developing a robust approach to using observed spectra ``en masse'' to derive photometric redshifts, and as only part of a larger, extended effort within the extragalactic astronomical community to prepare for future large-scale surveys. We hope that the introduction of SOMs as a form of (discrete) non-linear dimensionality reduction that can be used to enhance MCMC-based sampling approaches (or more generally to understand high-dimensional data and/or models) is useful to the community, and stress that there remains ample opportunities to significantly expand and improve upon the basic ideas outlined here. We explore several of these extensions in our companion paper Speagle \& Eisenstein (2015b), where we compare their performance against an idealized mock LSST+\textit{Euclid} catalog.

\section*{Acknowledgements}

JSS would like to thank Michael Brown, Peter Capak, and Daniel Masters for insightful discussions as well as Charles Alcock for supervising the senior thesis course where a portion of this work was completed. JSS is grateful for financial support from the CREST program, which is funded by the Japan Science and Technology (JST) Agency. This work has benefited extensively from access to Harvard University's Odyssey computing cluster, which is supported by the FAS Division of Science's Research Computing Group.

\appendix

\section{The Brown et al. (2014) Archetypes}
\label{ap:brown}

As discussed in \S\ref{subsec:phot_trad}, the most extensively used galaxy spectral energy distribution (SED) template libraries in the literature \citep[see, e.g.,][]{bolzonella+00,benitez00,feldmann+06,ilbert+13,carrascokindbrunner14b} are taken from \citet{coleman+80}, \citet{kinney+96}, and more recently \citet{polletta+07}. Although these templates have been exceptionally useful over the past several decades, they have understandable limitations: 
\begin{enumerate}
	\item Wavelength coverage is often limited to the ultraviolet and optical, which often necessitates the use of stellar population synthesis \citep{conroy13} models to extend them into the infrared.
	\item Many of the underlying spectra are often of galaxy \textit{nuclei}, rather than the entire integrated spectrum.
	\item Many of the templates predate the era of well-calibrated wide-field imaging, which are needed to guarantee accurate flux normalizations of the templates. Systematic errors that arise from this could significantly impact the photo-z fitting process.
	\item The color space coverage of these templates is incredibly sparse -- covered using only 4, 10, and 17 spectra for \citet{coleman+80}, \citet{kinney+96}, and \citet{polletta+07}, respectively -- and often does not track the observed locus of galaxy colors \citep{brown+14}.
\end{enumerate} 

To address these issues, \citet{brown+14} obtain an ``atlas'' of 129 SEDs of nearby ($z < 0.05$) galaxies with archival
drift-scan optical spectroscopy and mid-IR spectroscopy, all of which have well calibrated multi-wavelength imaging across 26 bands spanning from the UV to the mid-IR. Sample selection was set by the availability of suitable spectra and images, which fortunately results in a very diverse (but not necessarily representative) sample of 129 nearby galaxies. Our requirement for multi-wavelength images and spectroscopy results in our sample (largely) being a subset of other surveys of the nearby Universe. Our atlas spans a broad range of galaxy types, including ellipticals, spirals, merging galaxies, blue compact dwarfs, and luminous infrared galaxies.

Accurate matched-aperture photometry was obtained from archival imaging from the Galaxy Evolution Explorer \citep[GALEX;][]{morrissey+07}, Swift UV/optical monitor telescope \citep[UVOT;][]{roming+05}, Sloan Digital Sky Survey III \citep[SDSS-III][]{aihara+11}, Two Micron All Sky Survey \citep[2MASS;][]{skrutskie+06}, \textit{Spitzer Space Telescope} \citep{fazio+04,rieke+04} and Wide-field Infrared Space Explorer \citep[WISE;][]{wright+10}. The majority of the optical spectra are taken from \citet{moustakaskennicutt06}, \citet{moustakas+10}, \citet{gavazzi+04}, and \citet{kennicutt92}, while the mid-IR spectra are from \textit{Spitzer Space Telescope}'s Infrared Spectrograph \citep[IRS;][]{houck+04} and the \textit{Akari} infrared camera \citep[IRC;][]{onaka+07}. Gaps in spectral coverage were filled in using MAGPHYS \citep{dacunha+08} SED models which were carefully fit to the pre-existing spectrophotometry. This rich dataset of multi-wavelength spectrophotometric observations allowed the models to be normalized, constrained and verified to high accuracy.

\begin{figure*}
	\centering
	\includegraphics[scale=0.36]{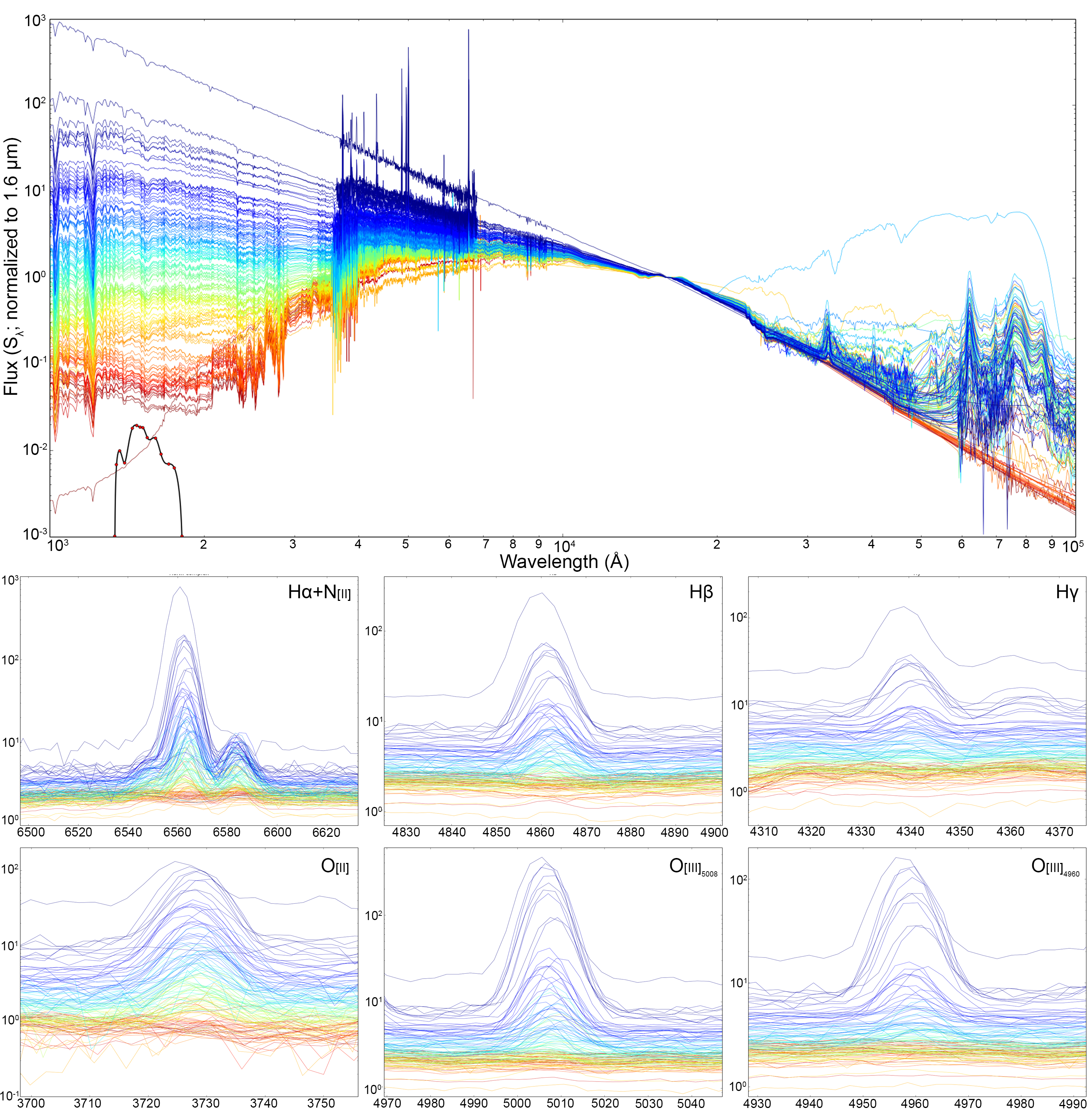}
	\caption{
		\textbf{Top panel:} The 129 \citet{brown+14} galaxy templates (normalized to 1.6\,\um), colored based on their \textit{GALEX} FUV flux (increasing from red to blue). A scaled version of the cubic spline fit (black) used to model the transmission curve (red circles) is shown for reference. Over wavelength ranges where a galaxy was not observed, the corresponding spectrum from \citet{brown+14} was derived by interpolating between the observed data using \texttt{MAGPHYS} \citep{dacunha+08}. Compared to previous template sets such as \citet{coleman+80}, \citet{kinney+96}, and \citet{polletta+07}, the \citet{brown+14} templates provide significantly denser coverage over a broader region of color space. \textbf{Bottom panel:} Insets showing spectral features around the central wavelengths for the H$\alpha$+N{\scriptsize[II]} complex (top left), H$\beta$ (top middle), H$\gamma$ (top right), O{\scriptsize[II]} (bottom left), O{\scriptsize[III]}$_{4960}$ (bottom middle), and O{\scriptsize[III]}$_{5008}$ (bottom right). The wide range in emission line strengths probed by the template set is easily visible.}
	\label{fig:brown_templates}
\end{figure*}

The 129 \citet{brown+14} spectral archetypes and zoom-in regions around their associated emission lines are shown in Figure~\ref{fig:brown_templates}. The corresponding set of emission line strengths (measured in terms of EW in {\AA}), normalized FUV fluxes, and other ancillary information is listed in Table~\ref{tab:brown_info}.

\onecolumn

\hfill \break


\begin{center}
	\begin{longtable}{l c c c c c c c c}
		\caption{Emission line strengths for the 129 \citet{brown+14} templates. Templates are ordered (1\,--\,129) by FUV flux in the \textit{GALEX} FUV filter after normalizing each template to $1.6$\,\um. Emission line strengths are measured using equivalent widths in \AA.}\\
		\hline
		\textbf{ID} & \textbf{Number} & \textbf{log FUV} & \textbf{H}$\mathbf{\alpha}$\textbf{+N{\scriptsize[II]}} & \textbf{H}$\mathbf{\beta}$ & \textbf{H}$\mathbf{\gamma}$ & \textbf{O{\scriptsize[II]}} & \textbf{O{\scriptsize[III]}}$_{\mathbf{4960}}$ & \textbf{O{\scriptsize[III]}}$_{\mathbf{5008}}$ \\
		\hline
		\endfirsthead
		\multicolumn{9}{c}%
		{\tablename\ \thetable\ -- \textit{Continued from previous page}} \\
		\hline
		\textbf{ID} & \textbf{Number} & \textbf{log FUV} & \textbf{H}$\mathbf{\alpha}$\textbf{+N{\scriptsize[II]}} & \textbf{H}$\mathbf{\beta}$ & \textbf{H}$\mathbf{\gamma}$ & \textbf{O{\scriptsize[II]}} & \textbf{O{\scriptsize[III]}}$_{\mathbf{4960}}$ & \textbf{O{\scriptsize[III]}}$_{\mathbf{5008}}$ \\
		\hline
		\endhead
		\hline \multicolumn{9}{c}{\textit{Continued on next page}} \\
		\endfoot
		\hline
		\endlastfoot
		IC 0860 & 1 & 0.107 & 2.093 & -2.809 & -0.545 & -0.985 & 1.22 & 3.395 \\
		NGC 4551 & 2 & 0.485 & -1.06 & -2.679 & 1.153 & -1.369 & 1.693 & 3.185 \\
		NGC 0584 & 3 & 0.517 & -2.305 & -3.275 & 1.367 & -0.991 & 1.194 & -1.896 \\
		NGC 4125 & 4 & 0.524 & 1.312 & -2.541 & 1.485 & -1.169 & 0.785 & 0.321 \\
		NGC 5866 & 5 & 0.629 & 0.504 & -2.226 & 1.535 & -1.494 & 1.269 & -0.225 \\
		IC 4051 & 6 & 0.653 & 0.839 & -2.837 & 0.935 & -1.361 & 2.335 & 2.303 \\
		NGC 4387 & 7 & 0.673 & -0.903 & -2.634 & 1.965 & -1.406 & 1.275 & 3.032 \\
		NGC 3190 & 8 & 0.69 & 0.743 & -3.291 & 0.625 & -0.903 & 1.306 & 1.838 \\
		NGC 3379 & 9 & 0.691 & -0.252 & -2.388 & 1.182 & -1.354 & 1.224 & -1.758 \\
		NGC 2388 & 10 & 0.697 & 30.736 & -1.367 & -1.924 & 0.187 & 1.713 & 8.677 \\
		NGC 4458 & 11 & 0.708 & -1.554 & -3.004 & 0.244 & -0.42 & 1.193 & -0.606 \\
		NGC 0474 & 12 & 0.715 & -0.654 & -2.643 & 0.723 & -0.918 & 1.425 & -2.591 \\
		NGC 4168 & 13 & 0.725 & 0.148 & -2.317 & 2.372 & -1.047 & 1.039 & -3.148 \\
		CGCG 049-057 & 14 & 0.751 & 13.56 & -4.241 & 0.899 & -1.143 & 4.896 & 9.134 \\
		NGC 4660 & 15 & 0.764 & -0.829 & -2.413 & 2.705 & -1.394 & 0.985 & 2.587 \\
		NGC 4594 & 16 & 0.791 & 0.509 & -1.691 & 3.09 & -0.346 & 1.879 & 8.15 \\
		Mrk 1490 & 17 & 0.803 & 44.245 & -1.905 & -1.965 & -0.145 & 1.057 & 4.119 \\
		NGC 4473 & 18 & 0.809 & 2.361 & 2.406 & 0.274 & 1.829 & -0.869 & -3.174 \\
		NGC 4621 & 19 & 0.809 & -0.456 & -2.681 & 1.311 & -1.873 & 1.501 & 3.142 \\
		NGC 7585 & 20 & 0.829 & 0.289 & -3.233 & 1.611 & -1.082 & 1.705 & 0.036 \\
		UGC 12150 & 21 & 0.838 & 21.282 & -2.907 & -2.154 & -0.022 & 1.54 & 10.977 \\
		NGC 4365 & 22 & 0.869 & -0.88 & -3.128 & 3.457 & -0.92 & 1.14 & 0.815 \\
		NGC 0750 & 23 & 0.876 & -0.002 & -2.313 & 2.388 & -1.319 & 1.104 & -2.772 \\
		NGC 4926 & 24 & 0.902 & -1.007 & -2.994 & 2.719 & -0.725 & 0.859 & -4.609 \\
		NGC 4550 & 25 & 0.948 & 0.681 & -2.788 & 0.249 & 0.244 & 1.678 & 2.25 \\
		NGC 4889 & 26 & 0.959 & -0.208 & -1.967 & 2.089 & -1.356 & 0.762 & -0.689 \\
		NGC 4552 & 27 & 0.98 & -0.092 & -2.167 & 2.788 & -1.141 & 1.156 & -1.498 \\
		NGC 5195 & 28 & 1.052 & 6.765 & -2.818 & -1.542 & -0.414 & 2.007 & 7.324 \\
		NGC 4450 & 29 & 1.129 & 3.2 & -2.456 & 2.121 & -0.768 & 1.398 & 3.853 \\
		NGC 4486 & 30 & 1.186 & 2.691 & -2.403 & 3.259 & -1.149 & 1.54 & -1.315 \\
		NGC 4860 & 31 & 1.187 & -3.396 & -2.544 & 1.558 & -0.888 & 0.78 & -0.608 \\
		IC 4553 & 32 & 1.234 & 16.539 & -4.391 & -3.407 & 0.62 & 1.153 & 8.988 \\
		Mrk 0331 & 33 & 1.303 & 56.039 & 0.701 & -2.108 & 2.226 & 1.085 & 10.099 \\
		NGC 7331 & 34 & 1.31 & 7.446 & -2.639 & 0.81 & -0.681 & 0.359 & 1.515 \\
		IRAS 17208-0014 & 35 & 1.313 & -0.212 & -1.327 & -4.747 & 0.317 & 1.456 & 20.681 \\
		NGC 0660 & 36 & 1.318 & 20.702 & -3.524 & -0.521 & 1.619 & 1.301 & 21.005 \\
		NGC 4725 & 37 & 1.338 & 4.78 & -1.411 & 0.711 & -1.317 & 0.674 & 1.759 \\
		NGC 5104 & 38 & 1.377 & 18.824 & -2.555 & -1.442 & 0.706 & 1.868 & 11.754 \\
		NGC 4579 & 39 & 1.391 & 7.225 & -2.305 & 0.848 & 0.029 & 1.433 & 7.788 \\
		UGC 05101 & 40 & 1.409 & 17.427 & -4.575 & -3.703 & 2.756 & 4.319 & 12.498 \\
		UGC 09618 N & 41 & 1.427 & 40.436 & 2.135 & 0.221 & 7.168 & 2.87 & 32.567 \\
		NGC 3521 & 42 & 1.459 & 13.709 & -2.11 & -0.347 & -0.196 & 1.481 & 3.727 \\
		NGC 4569 & 43 & 1.481 & 8.822 & -3.372 & -1.515 & -0.003 & 1.223 & 5.463 \\
		NGC 4826 & 44 & 1.5 & 9.252 & -1.557 & 0.865 & -0.922 & 1.571 & 3.146 \\
		CGCG 453-062 & 45 & 1.543 & 52.113 & 1.934 & -1.196 & 1.835 & 1.021 & 19.234 \\
		NGC 7771 & 46 & 1.606 & 22.34 & -1.452 & -0.786 & -0.129 & 1.346 & 10.524 \\
		NGC 5033 & 47 & 1.628 & 17.818 & -1.072 & 0.323 & 0.343 & 1.671 & 5.932 \\
		NGC 0520 & 48 & 1.658 & 6.313 & -3.395 & -3.124 & -0.495 & 0.829 & 5.008 \\
		NGC 6240 & 49 & 1.68 & 79.629 & 1.566 & -1.246 & 6.411 & 2.13 & 48.512 \\
		IC 5298 & 50 & 1.699 & 36.544 & -0.161 & -0.525 & 3.411 & 2.605 & 11.456 \\
		NGC 5953 & 51 & 1.707 & 50.435 & 1.817 & -0.583 & 4.405 & 3.181 & 10.896 \\
		Arp 118 & 52 & 1.741 & 25.38 & 0.126 & 1.11 & 2.583 & 2.095 & 9.063 \\
		NGC 4676 A & 53 & 1.752 & 14.152 & -2.084 & -2.041 & 1.402 & 1.526 & 4.635 \\
		UGC 08696 & 54 & 1.765 & 70.213 & 3.048 & -1.108 & 34.776 & 12.112 & 45.0 \\
		NGC 4138 & 55 & 1.766 & 10.713 & -1.264 & 0.441 & 0.902 & 1.584 & 11.997 \\
		NGC 2798 & 56 & 1.768 & 49.919 & 2.371 & 0.884 & 2.114 & 1.663 & 15.965 \\
		NGC 5055 & 57 & 1.784 & 12.001 & -1.078 & -3.394 & -0.356 & 1.28 & -0.372 \\
		NGC 3627 & 58 & 1.786 & 10.848 & -2.704 & -1.92 & -0.213 & 1.389 & 4.621 \\
		NGC 3351 & 59 & 1.786 & 13.65 & -0.605 & 1.313 & -0.387 & 1.45 & 2.904 \\
		UGC 04881 & 60 & 1.803 & 16.84 & -2.047 & -2.584 & 0.622 & 1.206 & 8.535 \\
		III Zw 035 & 61 & 1.808 & 29.747 & -0.877 & -2.309 & 6.639 & 2.309 & 22.798 \\
		NGC 1144 & 62 & 1.83 & 42.133 & 2.163 & 1.162 & 4.071 & 2.713 & 13.528 \\
		UGC 09618 & 63 & 1.836 & 42.771 & 2.2 & -0.387 & 6.339 & 2.255 & 23.991 \\
		NGC 7591 & 64 & 1.859 & 26.893 & -0.57 & -1.897 & 1.599 & 1.923 & 6.187 \\
		NGC 3079 & 65 & 1.944 & 32.715 & 0.525 & -0.06 & 4.996 & 2.519 & 23.723 \\
		IC 0883 & 66 & 1.981 & 23.415 & -3.82 & -4.609 & 1.509 & 1.326 & 11.704 \\
		NGC 1068 & 67 & 2.003 & 65.914 & 2.164 & 0.376 & 29.549 & 7.144 & 12.32 \\
		NGC 4536 & 68 & 2.012 & 31.491 & 0.9 & 0.826 & 0.224 & 1.318 & 15.582 \\
		NGC 3265 & 69 & 2.036 & 35.734 & 1.603 & 0.792 & 1.498 & 2.147 & 16.861 \\
		NGC 3198 & 70 & 2.057 & 17.223 & -0.859 & -0.399 & 1.951 & 1.807 & 13.162 \\
		NGC 5653 & 71 & 2.068 & 58.944 & 3.374 & -0.537 & 0.984 & 0.938 & 11.779 \\
		NGC 2623 & 72 & 2.079 & 9.589 & -4.921 & -4.404 & -0.126 & 1.142 & 7.916 \\
		NGC 5258 & 73 & 2.107 & 32.256 & 0.149 & -1.005 & -0.583 & 2.582 & 1.089 \\
		NGC 4088 & 74 & 2.115 & 33.83 & 1.631 & -1.534 & 2.359 & 1.336 & 13.105 \\
		NGC 0695 & 75 & 2.133 & 97.139 & 6.507 & -0.649 & 5.296 & 2.956 & 23.144 \\
		NGC 5256 & 76 & 2.138 & 110.947 & 12.058 & 4.841 & 60.713 & 20.182 & 85.526 \\
		NGC 4254 & 77 & 2.157 & 39.605 & 1.402 & -1.309 & -0.189 & 1.573 & 5.274 \\
		NGC 5194 & 78 & 2.217 & 26.874 & -0.614 & -1.637 & -0.333 & 1.263 & 0.721 \\
		NGC 5713 & 79 & 2.225 & 47.146 & 1.592 & -1.745 & 1.105 & 1.708 & 17.021 \\
		NGC 1614 & 80 & 2.237 & 104.227 & 4.821 & -1.102 & 6.261 & 2.538 & 17.564 \\
		NGC 4321 & 81 & 2.252 & 23.713 & -0.461 & -0.685 & -0.22 & 1.7 & 5.446 \\
		UGC 08335 NW & 82 & 2.261 & 32.343 & -1.747 & -0.257 & 6.528 & 1.71 & 18.187 \\
		UGC 09618 S & 83 & 2.293 & 49.65 & 2.635 & -0.873 & 5.875 & 1.522 & 19.294 \\
		NGC 3938 & 84 & 2.311 & 18.116 & 0.812 & -5.755 & -3.309 & 4.478 & -6.176 \\
		NGC 0855 & 85 & 2.318 & 24.483 & 1.49 & 0.489 & 9.411 & 3.923 & 35.78 \\
		NGC 4194 & 86 & 2.33 & 113.093 & 7.625 & 0.224 & 11.079 & 4.465 & 26.901 \\
		NGC 0628 & 87 & 2.333 & 6.353 & -2.941 & -1.163 & -1.667 & 0.672 & -1.063 \\
		UGC 08335 SE & 88 & 2.35 & 32.294 & -1.747 & -0.257 & 6.528 & 1.71 & 18.187 \\
		IC 0691 & 89 & 2.39 & 133.409 & 21.446 & 7.524 & 48.353 & 16.898 & 95.585 \\
		NGC 7674 & 90 & 2.403 & 64.005 & 4.323 & 1.334 & 32.929 & 4.971 & 16.675 \\
		IRAS 08572+3915 & 91 & 2.423 & -4.044 & -1.336 & -5.32 & 4.374 & 3.442 & 19.989 \\
		NGC 1275 & 92 & 2.431 & 52.819 & 2.496 & 0.632 & 11.547 & 0.866 & 40.914 \\
		NGC 4631 & 93 & 2.438 & 42.17 & 2.692 & -1.617 & 10.731 & 4.283 & 30.24 \\
		CGCG 436-030 & 94 & 2.452 & 79.679 & 4.41 & -1.041 & 6.004 & 2.232 & 32.848 \\
		NGC 4385 & 95 & 2.512 & 56.745 & 5.835 & -0.602 & 3.351 & 2.771 & 24.507 \\
		NGC 3690 & 96 & 2.529 & 172.191 & 18.153 & 6.217 & 24.43 & 8.209 & 53.27 \\
		NGC 4625 & 97 & 2.552 & 30.304 & 0.299 & -1.563 & 0.24 & 1.276 & 13.633 \\
		NGC 6090 & 98 & 2.564 & 174.871 & 16.692 & 3.879 & 13.042 & 4.88 & 32.205 \\
		NGC 7592 & 99 & 2.576 & 82.765 & 7.225 & 1.585 & 11.748 & 3.459 & 32.072 \\
		NGC 7679 & 100 & 2.599 & 96.514 & 6.597 & -0.721 & 15.22 & 4.445 & 27.065 \\
		NGC 5257 & 101 & 2.615 & 84.235 & 6.921 & 1.296 & 6.796 & 3.735 & 24.628 \\
		NGC 0337 & 102 & 2.649 & 56.846 & 4.958 & -0.917 & 11.962 & 5.438 & 36.423 \\
		NGC 4559 & 103 & 2.66 & 40.925 & 3.644 & 0.484 & 5.62 & 3.241 & 28.897 \\
		NGC 3049 & 104 & 2.682 & 65.395 & 6.997 & 2.529 & 2.955 & 2.176 & 21.121 \\
		Arp 256 S & 105 & 2.706 & 144.653 & 14.351 & 2.689 & 16.251 & 5.041 & 45.095 \\
		NGC 5992 & 106 & 2.722 & 69.875 & 3.065 & -2.129 & 8.017 & 3.329 & 24.366 \\
		NGC 2537 & 107 & 2.775 & 44.597 & 5.093 & 1.031 & 14.247 & 5.294 & 39.62 \\
		NGC 2403 & 108 & 2.796 & 40.263 & 2.204 & 0.348 & 5.364 & 3.589 & 19.226 \\
		Arp 256 N & 109 & 2.873 & 78.794 & 9.213 & 2.309 & 10.968 & 4.424 & 35.083 \\
		NGC 3870 & 110 & 2.902 & 50.72 & 4.045 & -0.236 & 15.81 & 5.847 & 51.733 \\
		NGC 7714 & 111 & 2.926 & 166.966 & 20.989 & 6.949 & 39.409 & 14.626 & 50.784 \\
		Mrk 33 & 112 & 2.962 & 136.25 & 19.54 & 6.214 & 42.64 & 15.514 & 66.661  \\
		UGC 06665 & 113 & 3.001 & 191.848 & 29.257 & 10.648 & 88.398 & 28.317 & 93.363 \\
		II Zw 096 & 114 & 3.015 & 130.53 & 14.586 & 3.57 & 32.748 & 11.695 & 51.419 \\
		UGCA 208 & 115 & 3.032 & 34.082 & 18.01 & 5.174 & 58.413 & 19.799 & 75.589 \\
		NGC 7673 & 116 & 3.064 & 121.199 & 15.499 & 2.906 & 34.784 & 12.212 & 56.955 \\
		NGC 6052 & 117 & 3.065 & 153.152 & 18.896 & 3.937 & 41.892 & 14.156 & 66.227 \\
		NGC 3773 & 118 & 3.071 & 75.971 & 12.059 & 4.264 & 20.9 & 6.446 & 47.118 \\
		NGC 3310 & 119 & 3.106 & 144.269 & 17.445 & 3.693 & 44.222 & 14.069 & 56.036 \\
		NGC 4670 & 120 & 3.135 & 105.84 & 18.313 & 5.693 & 50.831 & 17.596 & 53.653 \\
		UGCA 410 & 121 & 3.25 & 234.161 & 46.254 & 14.24 & 251.659 & 84.87 & 80.332 \\
		Mrk 0475 & 122 & 3.371 & 335.229 & 78.665 & 25.074 & 413.894 & 127.431 & 90.324 \\
		Haro 06 & 123 & 3.402 & 196.297 & 36.071 & 10.266 & 114.95 & 42.17 & 95.727 \\
		Mrk 1450 & 124 & 3.504 & 430.038 & 96.572 & 28.213 & 483.533 & 160.582 & 117.336 \\
		UM 461 & 125 & 3.509 & 364.163 & 70.45 & 19.215 & 390.496 & 130.945 & 0.955 \\
		Mrk 0930 & 126 & 3.555 & 410.372 & 69.668 & 22.229 & 308.53 & 96.711 & 113.184 \\
		UGC 06850 & 127 & 3.627 & 287.527 & 52.295 & 13.046 & 228.586 & 79.041 & 75.651 \\
		UGCA 219 & 128 & 3.827 & 148.074 & 24.499 & 5.611 & 100.754 & 30.881 & 40.005 \\
		UGCA 166 & 129 & 4.504 & 856.074 & 129.955 & 39.924 & 235.194 & 72.863 & 24.653
		\label{tab:brown_info}
	\end{longtable}
\end{center}

\twocolumn



\bibliography{photoz}


\bsp	
\label{lastpage}
\end{document}